\documentclass[journal]{IEEEtran}

\IEEEoverridecommandlockouts           
\usepackage{epsfig} 
\usepackage{epstopdf}
\usepackage{amsmath,amssymb,amsfonts}
\usepackage{color}
\usepackage{lipsum}
\usepackage{balance}
\usepackage{booktabs,tabularx,longtable,ifsym}
\usepackage{eurosym}
\usepackage{multirow}
\usepackage{mathtools}
\usepackage{pdfrender}
\usepackage{trfsigns}

\usepackage{centernot}
\usepackage{color}
\usepackage[acronym]{glossaries}
\usepackage[ruled]{algorithm2e}
\usepackage{graphicx}
\usepackage{wrapfig}
\usepackage{subfig}
\usepackage{url}

\graphicspath{{Images/}}

\newcommand{\N}{\mathbb{N}}

\newcommand{\mc}[1]{\mathcal{#1}}

\newcommand{\col}{\mathrm{col}}
\newcommand{\tup}[1]{\textup{#1}}
\newcommand{\bs}[1]{\boldsymbol{#1}}
\newcommand{\QEDopen}{\hfill $\square$}

\newtheorem{remark}{Remark}

\makeglossaries
\newacronym{MIP}{MIP}{Mixed-Integer Programming}
\newacronym{SoC}{SoC}{State of Charge}
\newacronym{PEV}{PEV}{Plug-in Electric Vehicle}
\newacronym{EV}{EV}{Electric Vehicle}
\newacronym{MLD}{MLD}{Mixed-Logical-Dynamical}
\newacronym{GS}{GS}{Gauss-Southwell}
\newacronym{GNEP}{GNEP}{Generalized Nash Equilibrium Problem}
\newacronym{MI-GPG}{MI-GPG}{Mixed-Integer Generalized Potential Game}
\newacronym{MINE}{$\varepsilon$-MINE}{$\varepsilon$-Mixed-Integer Nash Equilibrium}
\newacronym{CTM}{CTM}{Cell Transmission Model}
\newacronym{CS}{CS}{Charging Station}
\newacronym{FV}{FV}{fuel vehicle}
\newacronym{r2s}{r2s}{}
\newacronym{s2r}{s2r}{}
\newacronym{HO}{HO}{Highway Operator}
\newacronym{TDM}{TDM}{Traffic Demand Management}
\newacronym{ATDM}{ATDM}{Active Traffic Demand Management}
\newacronym{NDW}{NDW}{Nationaal
Dataportaal Wegverkeer}

\begin{document}
\title{Highway Traffic Control via Smart\\ e-Mobility -- Part II: Dutch A$13$ Case Study}

\author{Carlo Cenedese$^{\textrm{(a)}}$, Michele Cucuzzella $^{\textrm{(b),\,(c)}}$, Jacquelien M. A. Scherpen$^{\textrm{(c)}}$,\\ Sergio Grammatico$^{\textrm{(d)}}$ and Ming Cao$^{\textrm{(c)}}$
\thanks{$^{\textrm{(a)}}$ Department of Information Technology and Electrical Engineering, ETH Zürich, Zurich, Switzerland ({\texttt{ccenedese@ethz.ch}}).
$^{\textrm{(b)}}$ 
Department of Electrical, Computer and Biomedical Engineering, University of Pavia, Pavia, Italy ({\texttt{michele.cucuzzella@unipv.it}}).
$^{\textrm{(c)}}$ Jan C. Willems Center for Systems and Control, ENTEG, Faculty of Science and Engineering, University of Groningen, The Netherlands	({\texttt{\{j.m.a.scherpen, m.cao\}@rug.nl}}). 
$^{\textrm{(d)}}$ Delft Center for Systems and Control, TU Delft, The Netherlands
	({\texttt{s.grammatico@tudelft.nl}}). 	
	The work of Cenedese and Cao was supported by The Netherlands Organization for Scientific Research (NWO-vidi-14134), the one of Cucuzzella and Scherpen by the EU Project \lq MatchIT' (82203), and the one of Grammatico by NWO under project OMEGA (613.001.702) and P2P-TALES (647.003.003) and by the ERC under research project COSMOS (802348).}}

\markboth{IEEE - Transaction on Intelligent Transportation Systems, \today}%
{Shell \MakeLowercase{\textit{et al.}}: Bare Demo of IEEEtran.cls for IEEE Journals}

\maketitle

\begin{abstract}
In this paper, we study how to alleviate  highway traffic congestions by encouraging plug-in electric and hybrid vehicles to stop at  charging stations around peak congestion times. Specifically, we focus on a case study and simulate the adoption of a dynamic charging price depending on the traffic congestion. We use real traffic data of the A$13$ highway stretch between The Hague and Rotterdam, in The Netherlands, to identify the Cell Transmission Model. Then, we apply the algorithm proposed in \cite{cenedese:2020:highway_control_pI}(Part I: Theory) to different scenarios, validating the theoretical results and showing the benefits of our strategy in terms of traffic congestion alleviation. Finally, we carry out a sensitivity analysis of the proposed algorithm and discuss how to optimize its performance.
\end{abstract}


%
%
\IEEEpeerreviewmaketitle
\section{Introduction}
\label{sec:intro}
\IEEEPARstart{W}{e} study how to alleviate highway traffic congestions by encouraging plug-in hybrid and electric  vehicles to stop at charging stations around peak congestion time. In the first part of our work (Part I: Theory)~\cite{cenedese:2020:highway_control_pI}, motivated by the rising number of \glspl{PEV} and inspired by the conventional ramp metering control, we have proposed a novel \gls{ATDM} strategy based on soft measures to alleviate traffic congestion during rush hours. Specifically, we have designed a pricing policy to make the charging price dynamic and dependent on the (predicted) traffic congestion level, encouraging the \gls{PEV} owners to stop for charging when the congestion level is (going to be) high. To achieve this goal, we have developed a novel framework that models how the proposed policy affects the drivers' decisions as a \gls{MI-GPG}. From a technological point of view, we have introduced the concept of ``road-to-station'' \gls{r2s} and ``station-to-road'' \gls{s2r} traffic flows, and  shown that the selfish behavior of the drivers leads to decision strategies that are individually optimal in the sense of Nash.

In this paper (Part II: Case Study), the proposed \gls{ATDM} strategy is validated in simulation on the A$13$ highway stretch between The Hague and Rotterdam, in The Netherlands.
More precisely, the main contribution of this paper is to show by simulation that the adoption of a dynamic charging price policy depending on the traffic congestion can alleviate highway traffic congestion by encouraging plug-in electric and hybrid vehicles to stop at one \gls{CS} around peak congestion time. To the best of the authors' knowledge, this is the first study that proposes the \gls{r2s} and \gls{s2r} technology to develop an \gls{ATDM}.
Furthermore, to support the reliability of our simulation results, we use real traffic data to identify the parameters of the corresponding \gls{CTM}, which predicts the evolution of the traffic congestion~\cite[Sec.~II]{cenedese:2020:highway_control_pI}. Finally,  to show the general validity of the proposed approach and  to study its performance in different scenarios, we perform an extensive sensitivity analysis on the  system parameters, which allows us to identify the ``best'' setup that maximizes the performance of the proposed strategy, based on empirical results.

%
\textit{Structure of the paper:} In Section~\ref{sec:revisit_model_ptI}, we recall the most important components of the \gls{ATDM} strategy developed in the first part of this work (Part I: Theory)~\cite{cenedese:2020:highway_control_pI} and the associated notation. In Section~\ref{sec:CTM_ident}, we use real data to identify the parameters of the  \gls{CTM} and discuss the choice of the parameters adopted in  the decision making process. The first simulation is analyzed in Section~\ref{sec:simulation_section}, where the parameters reflect the current market scenario. For the most relevant decision parameters, a sensitivity analysis is carried out in Section~\ref{sec:sensitivity}.  Finally, Sections~\ref{sec:policies} and \ref{sec:conclusion}, respectively, discuss the policies that emerge  from the simulation results as the most effective and the conclusions that summarize the strengths of our approach together with  promising future developments. 
\section{The \gls{ATDM} strategy proposed in Part~I}
\label{sec:revisit_model_ptI}
In this section,  we briefly review the \gls{ATDM} strategy we have proposed in the first part of this work (Part I: Theory) \cite{cenedese:2020:highway_control_pI}, including the problem formulation, theoretical background and notations (see Table~\ref{tab:notation}).

The traffic evolution is modeled via the \gls{CTM} that is a simple, yet sufficently accurate, first order model for traffic evolution on a highway~\cite{rinaldi:ferrara:2012:CTM_identification}. We consider a highway stretch in which there are no entering or exiting ramps, divided in $N$ cells of length $L_\ell$ where $\ell\in\mc N\coloneqq \{1,\dots,N\}$. Each cell $\ell$ has an entering flow of vehicles $\phi_\ell$  and exiting one $\phi_{\ell+1}$. We assume that the \glspl{PEV} leaving cell $1$ can choose to stop at the unique \gls{CS}, that we assume  to be  located between cells $1$ and $2$ (see \cite[Fig.~1]{cenedese:2020:highway_control_pI}). The flows of \glspl{PEV} entering and exiting the \gls{CS} are  \gls{r2s} and \gls{s2r} respectively. The relation between a cell's flow $\phi_\ell$ and its density $\rho_\ell$ describes the traffic evolution as defined  in~\cite[Eq.1--4]{cenedese:2020:highway_control_pI}.

We consider that the \gls{HO} regulates the price of the electricity purchased by the \glspl{PEV} at the \gls{CS}. This price $p$ is dynamic and subject to a discount that is proportional to the level of the actual (or predicted) traffic congestion: 
\smallskip
\begin{equation}
\label{eq:price}
\forall k\in\N \,,\: p(k) \coloneqq c_1 \: d(k) + c_2  u^{\textup{PEV}}(k) - c_3\sum_{\ell=2}^N \Delta_\ell(k)  \,,
\end{equation}
where $c_1,c_2, c_3>0$ and $\Delta_\ell$ is the cell's extra travel time due to the traffic congestion.
\begin{table}[t]
  \centering
  \caption{The variables and parameters (respectively in the upper and bottom part of each table) related to the \gls{CTM} and the charging scheduling decision problem.}
\label{tab:notation}
  \begin{tabular}{c c l}
  \toprule
\multicolumn{3}{c}{\gls{CTM} cell $\ell\in\mc N$}\\
\midrule
$\phi_\ell$&$[\textup{veh/h}]$& flow entering the cell\\
\gls{r2s}&$[\textup{veh/h}]$& flow entering the \gls{CS} \\
\gls{s2r}&$[\textup{veh/h}]$& flow exiting the \gls{CS}\\
$	\rho_\ell$ & $[\textup{veh/km}]$ & traffic density \\
$\Delta_\ell$ & $[\tup h]$ & extra travel time due to congestion\\
$\hat \Delta_\ell$ & $[\tup h]$ & predicted future extra travel time\\ & & due to congestion\\
\midrule
$L_\ell$ & $[\tup{km}]$&  cell length\\ 
$\bar v_\ell$ & $[\tup{km/h}]$&   free-flow velocity\\ 
$w_\ell$ & $[\tup{km/h}]$ & congestion wave speed\\
$q_\ell^{\max}$&$[\tup{veh/h}]$& maximum cell capacity\\
$\rho_\ell^{\max}$&$[\tup{veh/km}]$& maximum jam density\\
$T$ & $[\tup h]$ & length of the time interval\\
\bottomrule \\
\toprule
\multicolumn{3}{c}{Decision making process of agent $i\in\mc I$}\\
\midrule
$x_i$ & -- & battery \gls{SoC} of \gls{PEV} $i$\\
$u_i$ & $[\tup{kWh}]$ & energy purchased by \gls{PEV} $i$\\
$\delta_i$ & -- & binary var. $1$ iff \gls{PEV} $i$ is charging\\
$u^{\tup{PEV}}$& $[\tup{kWh}]$ & total energy purchased by \glspl{PEV}\\
$p$ & $[$\euro$\tup{/kWh}]$ &  price for $1\tup{kWh}$ at \gls{CS} \\
$\hat p$ & $[$\euro$\tup{/kWh}]$ &  predicted future price at \gls{CS} \\
$t_i$ & $[\tup h]$ & interval in which \gls{PEV} $i$ enters cell $2$ \\
$\vartheta_i$ & -- & rectangular function centered in $t_i$ \\
$\xi$ & $[\tup h]$  & (over)estimation of the extra travel time\\ && due to congestion\\
$\xi^{\tup{CS}}_i$ & $[\tup h]$  & estimation of the extra travel time\\ && due to \glspl{PEV} exiting the \gls{CS} \\
\midrule
$x_i^{\tup{ref}}$ & -- & min. \gls{SoC} of \gls{PEV} $i$ before leaving the \gls{CS}\\
$p_{\tup{EV}}$ & -- & percentage of \glspl{PEV} in the market\\ 
$lT$ & $[h]$ & length of the time interval\\
$T_{\tup h}$ & -- & number of intervals composing the  horizon \\
$C_i$ & $[\tup{kWh}]$  & battery capacity\\
$\eta_i$ & $[\tup{1/kWh}]$ & scaling associated to battery efficiency\\
$\alpha_i$ & -- & interest of \gls{PEV} $i$ in saving money\\
$\bar \delta$ & -- & number of charging spots at the \gls{CS}\\
$u^{\tup{max}}$ &$[\tup{kWh}]$  & \gls{CS} max. energy providable in 	an interval\\
$\overline u (\underline u)$& $[\tup{kWh}]$ & upper(lower) bound on purchasable energy\\
$d$ & $[\tup{kWh}]$ & base energy demand in the grid\\
$c_1,c_2$ & $[$\euro$\tup{/kWh}^2]$ & price scaling factors\\
$c_3$ & $[$\euro$\tup{/(h}\cdot\tup{kWh)}]$ & price scaling factor\\
$\overline p_i$ & $[$\euro$\tup{/kWh}]$ & fixed avg. price for $1\tup{kWh}$ not at \gls{CS}\\
$\beta_0$ & $[$\euro$\tup{/kWh}]$ & estimated price scaling factor\\
$\beta_1$ & $[$\euro$\tup{/(h}\cdot\tup{kWh)}]$ & estimated price scaling factor\\
$W$ & -- & half width of $\vartheta_i$ \\  
$\gamma$ & $[\tup h]$ & scaling used to compute in $\xi^{\tup{CS}}_i$ \\
$\chi$ & $[\tup h]$ & time-varying scaling of the travel time  \\
$\upsilon$ & -- & weight over the time spent at the \gls{CS}  \\
\bottomrule
\end{tabular}
\end{table}
The \glspl{PEV} exiting cell $1$ face a choice, namely, whether or not it is convenient to stop at the \gls{CS} given their \gls{SoC} $x_i$, the available plugs $\bar \delta$ and the current/predicted traffic situation. Sharing the facility and the dependency of $p$ in \eqref{eq:price} on $u^{\tup{PEV}}$ make the choice of the optimal charging strategy a challenging problem that has to be solved locally by each \gls{PEV}. Each driver aims at finding the best  trade-off between the money saved by exploiting the electricity price discount and the extra travel time due to the stop at the \gls{CS}.  This is modeled via the following two-term cost function: 
\begin{equation}\label{eq:Ji}
J_i\coloneqq \alpha_i J_i^{\tup{price}} + (1-\alpha_i) J_i^{\tup{time}},
\end{equation} 
which each \gls{PEV} (or agent) $i\in\mc I(k)$, with $\mc I(k)$ being the set of all the agents involved in the game, desires to minimize. The coefficient $\alpha_i\in (0,1)$ is the realization of a Gaussian random variable, with mean $\mu_\alpha$ and variance $\sigma_\alpha$. It represents the interest of agent $i$ in saving money rather than time. The first term in \eqref{eq:Ji} describes the total savings, i.e.,
\begin{equation}
\label{eq:J_price}
J^{\tup{price}}_{i}(k) \coloneqq \sum_{t\in\mc T(k)} (\hat p(t)-\bar p_i)u_i(t)\:,
\end{equation}
 where $\mc T(k)\coloneqq\{k,k+1, \dots,k+T_{\tup h}\}$ is the prediction horizon. The predicted price $\hat p(t)$ that the \gls{HO} applies during $t\in\mc T(k)$ is computed by the following approximation: 
\smallskip
\begin{equation}
\label{eq:price_hat}
\hat p(t) \coloneqq c_1 \: d(t) - \left[\beta_0(t) +\beta_1(t)  \sum_{\ell=2}^N \hat \Delta_\ell(t)  \right]\,.
\end{equation}
To characterize the interval $t_i\in\mc T(k)$ during which agent $i$ enters cell $2$, we introduced in \cite[Sec.~III.A.3]{cenedese:2020:highway_control_pI} the rectangular function $\vartheta_i$ that takes non-zero values during the $W$ intervals preceding and succeeding $t_i$.  
The second term in \eqref{eq:Ji} models the  extra travel time that \gls{PEV} $i$ experiences for stopping at the \gls{CS} and/or for the  traffic situation, i.e.,  
\smallskip
\begin{equation}
\label{eq:J_time}
J^{\tup{time}}_{i}(k) \coloneqq \sum_{t\in\mc T(k)} \chi(t) \big[(t-k)\upsilon +
 \xi(t)+\xi^{\tup{CS}}_i(t)\big]\vartheta_i(t) . 
\end{equation} 
If agent $i$ enters cell $2$ during $t\in\mc T$,  $\xi(t)+\xi^{\tup{CS}}_i(t)$ describes the estimated remaining travel time to complete the transit through cells $\{2,\dots,N\}$. The value of $\xi(t)$ is computed by exploiting the \gls{CTM} as an oracle to predict the evolution of the traffic congestion (see \cite[Sec.~III.A.3]{cenedese:2020:highway_control_pI} for further details).

The resources available at the \gls{CS} are finite and must be shared among the \glspl{PEV}. This creates a coupling between the agents' decisions, both in the cost \eqref{eq:J_time} and in the constraints.
Next, we qualitatively list the most important constraints that \gls{PEV} $i\in\mc I(k)$ has to meet during the choice of the optimal feasible charging schedule. For clarity we refer the reader to the corresponding formal definitions in \cite[Sec.~III.B]{cenedese:2020:highway_control_pI}:
\begin{itemize}
\item Charging only if $i$ is at the \gls{CS}, \cite[Eq.~16]{cenedese:2020:highway_control_pI},
\item exit the \gls{CS} if $i$ stops charging, \cite[Eq.~17]{cenedese:2020:highway_control_pI},
\item if $i$ stops, it has to remain at the \gls{CS} for at least $2W+1$ intervals, \cite[Eq.~19]{cenedese:2020:highway_control_pI},
\item exit the \gls{CS} only if $x_i>x_i^{\tup{ref}}$, \cite[Eq.~20]{cenedese:2020:highway_control_pI},
\item total energy purchased by the \glspl{PEV} has to be lower than $u^{\tup{max}}$, \cite[Eq.~21]{cenedese:2020:highway_control_pI},
\item the number of \glspl{PEV} simultaneously charging must be lower than $\bar \delta$, \cite[Eq.~21]{cenedese:2020:highway_control_pI}.
\end{itemize}
All the constraints are cast in an affine form via auxiliary variables and described in a compact form by means of the matrices $A$ and $b$. Organizing all the decision variables associated to each \gls{PEV} $i$ in a vector $z_i$ (and $\bs z \coloneqq \col((z_i)_{i\in\mc I(k)})$) allows us to cast the charging problem above as an exact  \gls{MI-GPG}, defined via the following set of interdependent optimization problems
\begin{equation}
\label{eq:game_compact}
\forall i \in \mc I(k) \::\quad \min_{z_i\in\mc Z_i(k)} J_i(z_i,\bs z_{-i}| k) \quad \text{s.t.} \: A\bs z\leq b \,
\end{equation}  
    where $\mc Z_i(k)$ is the time-dependent set of feasible decisions of each agent $i$, \cite[Eq.~24]{cenedese:2020:highway_control_pI}. Algorithm $1$ in \cite{cenedese:2020:highway_control_pI} is based on the sequential mixed-integer best response. It ensures convergence to an \gls{MINE}, since \eqref{eq:game_compact} is an exact potential game \cite[Th.~1]{cenedese:2020:highway_control_pI} . The remainder of the paper is devoted to the  implementation of the \gls{ATDM} strategy proposed in (Part I: Theory) \cite{cenedese:2020:highway_control_pI}, and to show via numerical simulations its beneficial effects on traffic alleviation.  Moreover, we analyze how different configurations of the parameters in Table~\ref{tab:notation} affect the travel time.  
\section{\gls{CTM} identification}
\label{sec:CTM_ident}
\begin{figure}[t]
\centering
\includegraphics[width=\columnwidth]{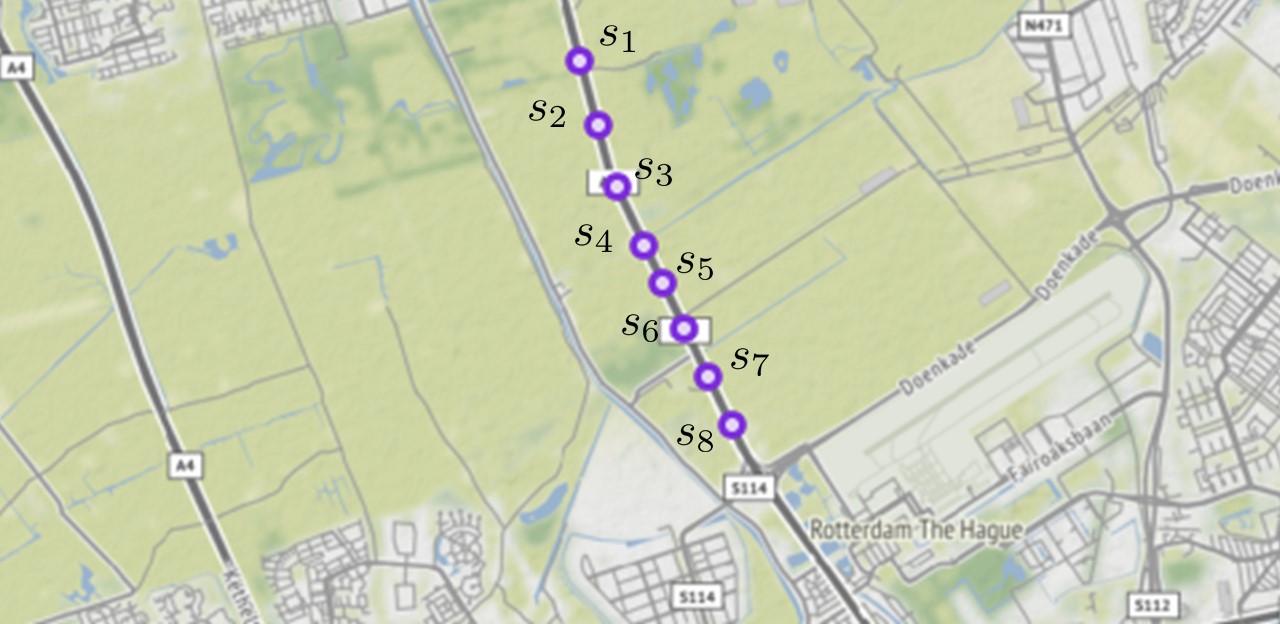}
\caption{Location along the A$13$ highway of the $8$ sensors selected for the simulations; \cite{ndw:dexter} indicates how to access this info.}\label{fig:sensor_location}
\end{figure}

\begin{figure}[t]
\centering
\includegraphics[width=\columnwidth]{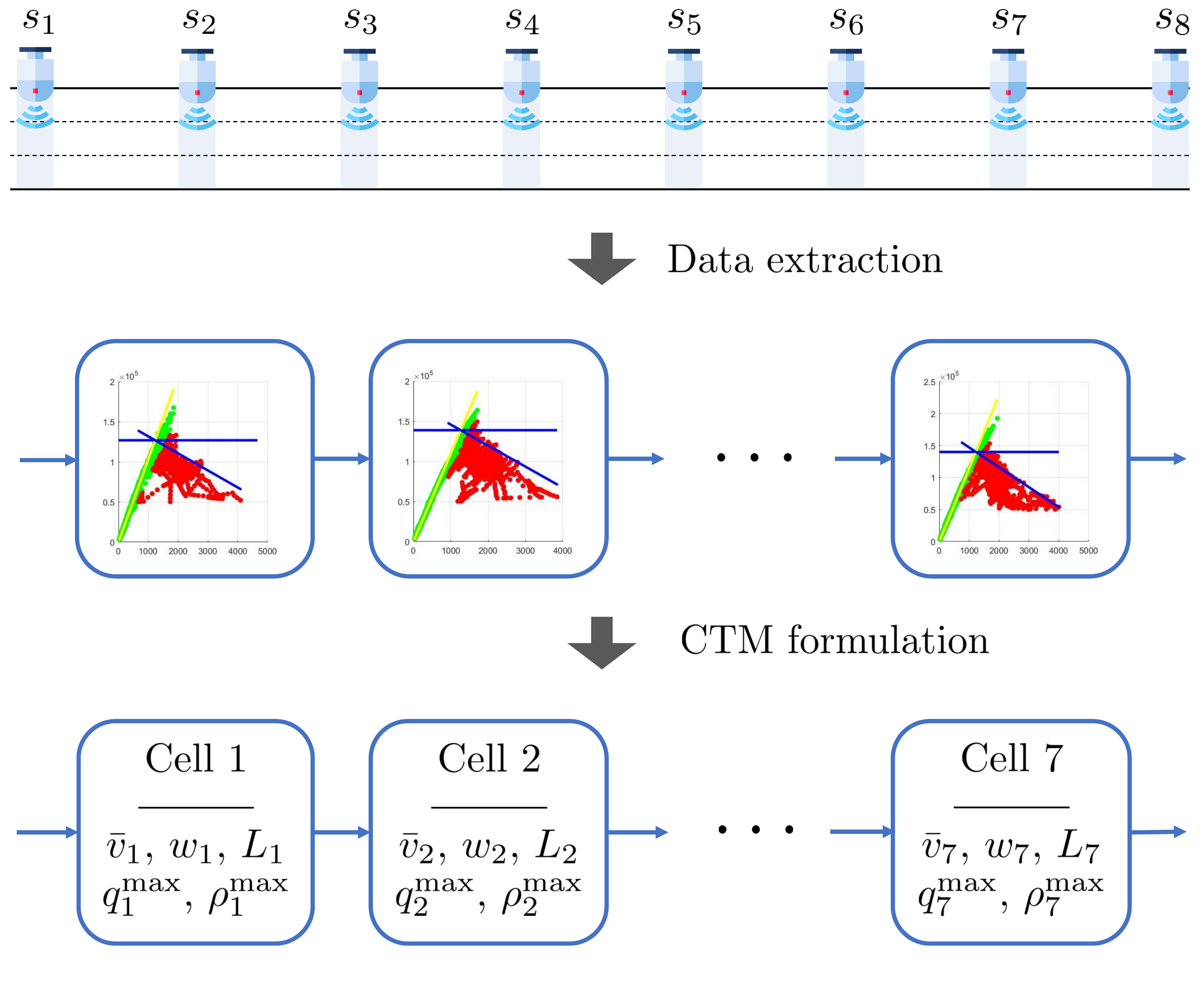}
\caption{Schematic structure of the procedure adopted to identify the parameters of the \gls{CTM}. }\label{fig:CTM_identification_flow}
\end{figure}

\begin{figure}[t]
\centering
    \subfloat[\label{subfig:fundamental}]{{\includegraphics[scale=0.9,width=\columnwidth]{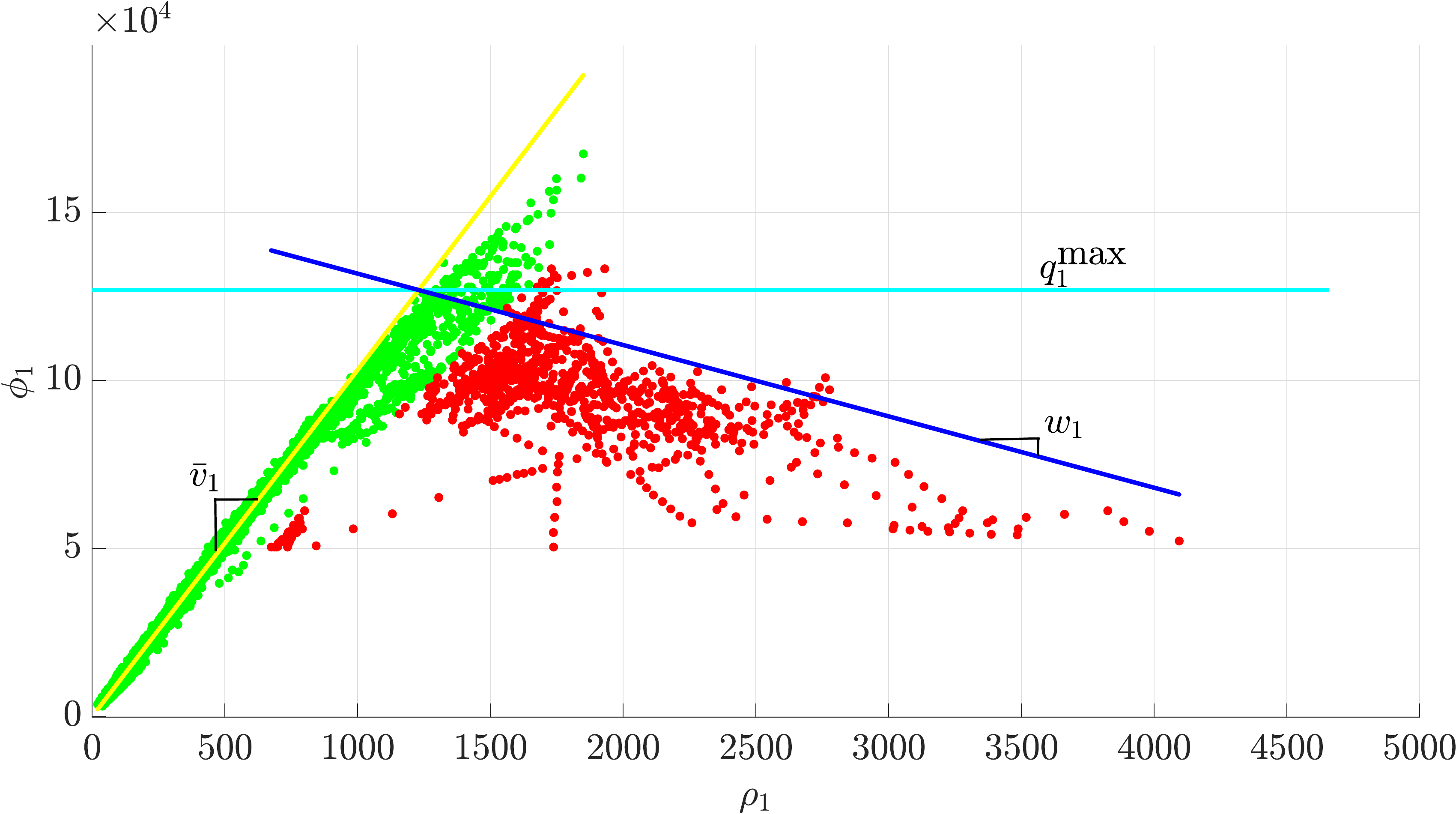}}}\\%
    \subfloat[\label{subfig:vel}]{{\includegraphics[scale=0.9,trim=00 0 00 0,clip=true, width=\columnwidth]{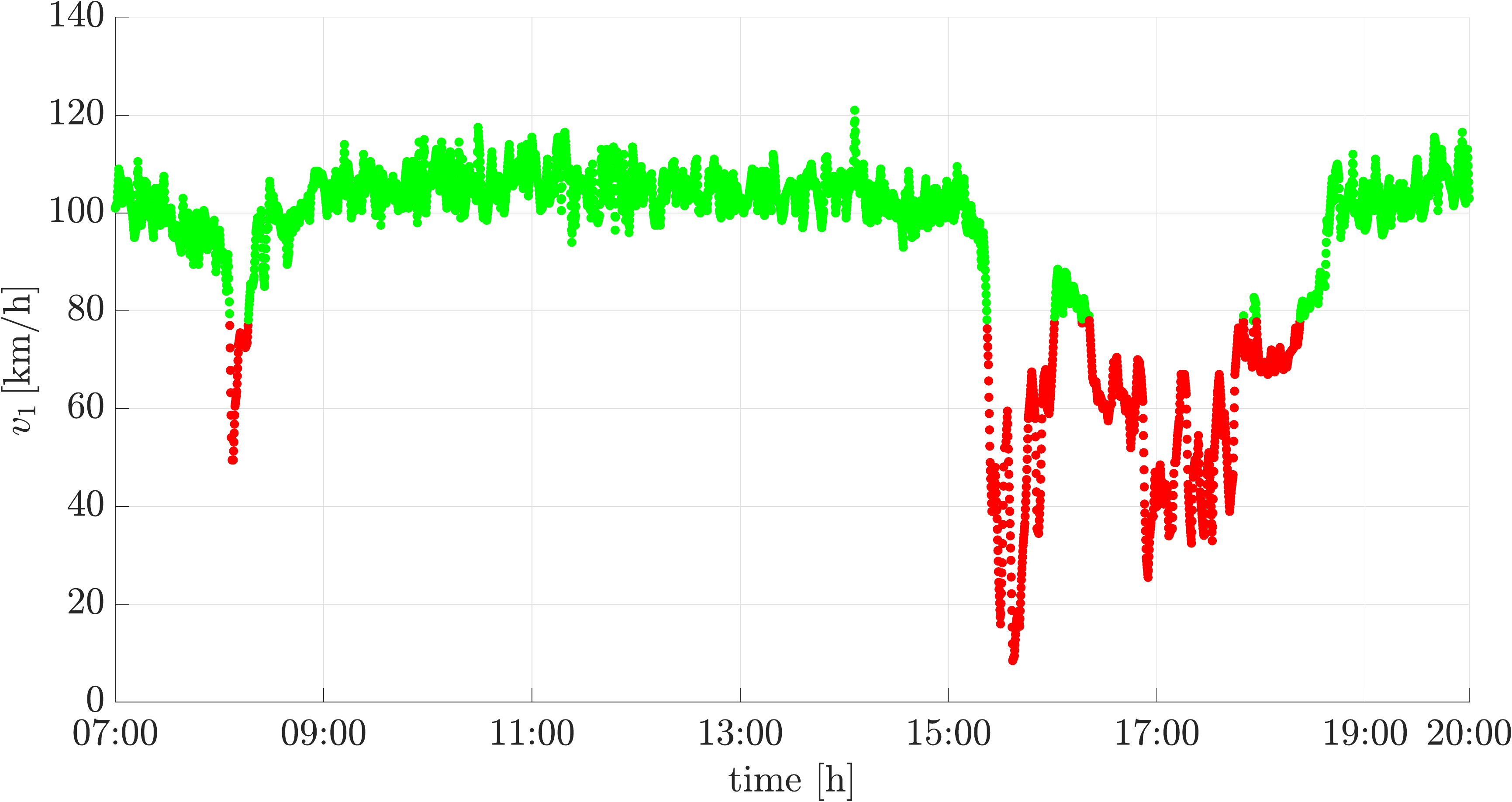} }}%
 \caption{(a) The fundamental diagram of cell $1$ obtained from the data of $s_1$. The data points are green if the cell is in a free-flow condition and red otherwise. (b) Average vehicles speed in cell $1$ during the day (the color legend is as in (a)).}
\label{fig:fundamental_&_velocity}
\end{figure}

Let us introduce the examined case study and identify the parameters of the associated \gls{CTM} starting from real-world data and following the schematic procedure depicted in Figure~\ref{fig:CTM_identification_flow}. We consider a stretch of the A$13$ highway in The Netherlands connecting the cities of The Hague and Rotterdam. The Dutch government embraces the policy of making traffic data  accessible via the \gls{NDW} \cite{ndw:portal}. In particular, the \gls{NDW}-Dexter portal \cite{ndw:dexter} allows us to access the sensors placed along the highway and extract the traffic data related to a particular date and time. Here, we select $8$ sensors, denoted in the following by $s_1,\dots,s_8$, on the roadway that from The Hague enters Rotterdam (Figure~\ref{fig:sensor_location}). 
The sensors are placed approximately at $500\,\tup m$ from each other  and measure  the number of vehicles and their average speed during a period of $1$ minute. We consider the data of the  $15/10/2019$ from $07$:$00$ to $20$:$00$, since there is no traffic congestion during the remaining hours.  
\smallskip
\begin{remark}[Minimum sample frequency]
The frequency of the data samples has to be high enough to allow the \gls{CTM} to evolve correctly. In fact, if for some $\ell\in\mc N$ the inequality  
$$\frac{T\bar v_\ell}{L_\ell}<1$$
is not satisfied, then the associated cell's density $\rho_\ell$ can take negative values. This is a consequence of the density definition  in \cite[Eq.~1]{cenedese:2020:highway_control_pI} and the lower bound for the sampling frequency is obtained from the worst-case scenario. One can solve this issue by considering longer cells or by interpolating the data obtained from the sensors to decrease the sampling time. For the problem at hand, we show that the free flow speed is approximately $120 \, \tup{km/h}$, therefore we interpolated  the data to obtain  a sampling time of $T=10\tup s$, so $T\bar v_\ell/L_\ell\simeq 0.66<1$.\hfill\QEDopen
\end{remark}
\smallskip
We construct a \gls{CTM} composed of $N=7$ cells, in which the flows $\phi_1,\dots,\phi_8$ between cells  are obtained directly from the raw data provided by the sensors, i.e., for all $j\in\{1,\dots,8\}$ 
$$ \phi_j\coloneqq \frac{\tup{\# veh. from }s_j}{T}.$$
 The the average speed of the vehicles is also measured by each sensor, thus for each cell $\ell\in\{1,\dots,7\}$  the density  is 
$$ \rho_\ell\coloneqq \frac{\phi_\ell}{\tup{avg. speed from }s_\ell}.$$
The relation between the flow and the density is represented for cell $1$ in the \textit{fundamental diagram} depicted in Figure~\ref{subfig:fundamental}, where we used a threshold on the average vehicles' velocity to distinguish between the data associated with the free flow and those measured during a traffic congestion. As in \cite{ferrara2018freeway} and more in details in \cite[Sec.~3.2]{li:2014:traffic_parameters}, we identified the parameters of each cell of the \gls{CTM} by using linear regression. For a cell $\ell$, the free flow speed $\bar v_\ell$ is  the slope of the linear regression applied to the data measured  during the free flow (the yellow line in Figure~\ref{subfig:fundamental}). Similarly, the congestion wave speed $w_\ell$ is the slope of the quantile regression applied to the leftover data. The result of the regression is illustrated via the blue line in Figure~\ref{subfig:fundamental}, and the density at which it intersects the horizontal axis represents the (theoretical) cell's maximum density $\rho_\ell^{\tup{max}}$. Finally, the intersection of the two regression lines (i.e., the blue and the yellow) defines the maximum cell capacity $q_\ell^{\tup{max}}$. Therefore, by applying these steps, sketched in Figure~\ref{fig:CTM_identification_flow} for all the cells, we  identify the parameters that, together with $\phi_1$ and $\phi_8$, fully describe the \gls{CTM}. Their final values are reported in Table~\ref{tab:CTM} and the code used is available at \cite{cenedese:github_ctm}.
\begin{table*}[t]
  \centering
  \caption{The variables and parameters (respectively in the upper and bottom part of each table) related to the \gls{CTM} and the charging scheduling decision problem.}
\label{tab:CTM}
  \begin{tabular}{l  c | c c c c c c c }
\toprule
  \multicolumn{2}{c}{Parameters} &
\multicolumn{7}{c}{Identified  values of the \gls{CTM} }\\
\midrule
$\ell$ &--& $1$ & $2$ & $3$ & $4$ & $5$ & $6$ & $7$\\
$L_\ell$ & $[\tup{km}]$&  $0.39$ & $0.41$ & $0.365$ & $0.365$ & $0.365$ & $0.5$ & $0.5$\\
$T$ & $[\tup{s}]$&   $10$ & $10$  & $10$  & $10$  & $10$  & $10$  & $10$ \\ 
$\bar v_\ell$ & $[\tup{km/h}]$ &   $103.15$ & $109.34$  & $111.67$  & $112.77$  & $113.07$  & $114.12$  & $114.18$ \\ 
$w_\ell$ & $[\tup{km/h}]$ & $21.23$ & $26.19$  & $19.91$  & $26.13$  & $27.73$  & $26.62$  & $31.11$ \\ 
$q_\ell^{\max}$&$[\tup{veh/h}]$& $1.26\cdot 10^{5}$ & $1.38\cdot 10^{5}$  & $1.28\cdot 10^{5}$  & $1.40\cdot 10^{5}$  & $1.40\cdot 10^{5}$  & $1.38\cdot 10^{5}$  & $1.40\cdot 10^{5}$ \\ 
$\rho_\ell^{\max}$&$[\tup{veh/km}]$& $7.2\cdot 10^{3}$ & $6.56\cdot 10^{3}$  & $7.58\cdot 10^{3}$  & $6.62\cdot 10^{3}$  & $6.29\cdot 10^{3}$  & $6.41\cdot 10^{3}$  & $5.73\cdot 10^{3}$ \\ 
\bottomrule
\end{tabular}
\end{table*}
\section{Decision process' parameters selection}
\label{sec:decision_proc_param}
In this section, we focus on the parameters describing the decision making process (or game) in which the \glspl{PEV} are involved during every time interval (see Table~\ref{tab:notation} for the description of each parameter). The chosen values are reported in Table~\ref{tab:C0_values} and characterize what we call the \textit{base case} in the remaining of the paper. Now, we list the motivations behind the choice of the values of the most significant parameters in the game.
\begin{itemize}
\item $p_{\tup{EV}}$: the current market share of \glspl{PEV} in the Netherlands is around $2\%$. However, \glspl{PEV} are expected to constitute more than $7\%$ of  total vehicles in the market by $2030$~\cite{iea:2020:ev_report}. For this reason, we choose  $p_{\tup{EV}}=5\%$. 
\item $l$: it is natural to consider the interval during which each player does not vary its strategy (namely if/when and for how long to stop at the \gls{CS}) longer than $T$. Then, we choose $lT=100\tup s$.  
\item $T_{\tup h}$: in \cite{verhoef:2010:incentive_traffic_managment} a group of commuters traveling via the A$12$ in the Netherlands participated in a test, whose results indicate that $60\%$ of the involved commuters could start their work within $30 \, \tup{min}$ later than usual.  Therefore, we considered $T_{\tup h}=15$, implying that the maximum time \glspl{PEV} can spend at the \gls{CS} (i.e., the length of the prediction horizon) is equal to $25\,\tup{min}$. 
\item $\bar\delta$: in The Netherlands, the government is putting a lot of effort in increasing the number of available charging points, doubling it between $2015$ and $2019$ \cite{rvo_chargign_spots}. Thus, we choose $\bar \delta=100$ to model the foreseeable increment in charging spots.   
\item $C_i$:  the battery capacity of each \gls{PEV} is randomly chosen from a pool of the most common \gls{PEV} models. Specifically, $C_i\in[12,100]$, where $12\,\tup{kWh}$ represents the capacity of a Mitsubishi Outlander and $100\,\tup{kWh}$ the one of a Tesla Model S.
\item $\alpha_i$: for the realization of this Gaussian variable we choose mean and variance equal to $\mu_\alpha=0.05$ and $\sigma_\alpha=0.03$, respectively. This choice reflects the assumption that agents are more keen to save time rather than money. The small variance mimics an homogeneous interest among the \glspl{PEV} owners. The case of a more heterogeneous population is analyzed in Section~\ref{sec:sensitivity}.
\item $\bar u$: the charging spots are able to fast charge the \glspl{PEV} and  the maximum energy that can be delivered by each spot in an hour is $150\,\tup{kWh}$. Thus, during a single time interval we have $\bar u =150\cdot lT = 4.16 \,\tup{kWh}$.
\item $u^{\max}$: the \gls{CS} is assumed to be able to deliver the maximum power to all the charging spots simultaneously.   
\item $\bar p_i$: this energy price is chosen equal to $0.205$\,\euro$\tup{/kWh}$ and represents the  average between the price for fast charging (see \cite{tesla_charger_price,fastnet}) and the one for  charging at home (see~\cite{iea:2020:energy_price_nl}). 
\item $d$: the base demand profile of the grid is obtained from \cite{eia_beta} and scaled to match the average demand of a regional grid in the Netherlands, so $d(k)\in[3.94, 4.16]\,\tup{MWh}$.
\item $c_1,c_2,c_3$: these values must be chosen simultaneously to make the average value of $p(t)$ equal to $\bar p_i$ and normalize the three components in the price. Furthermore, $c_3$ is particularly important, since it defines the discount due to the   traffic congestion. We introduce here what we refer to as  \textit{incentive} in the remainder of the paper. We say that we apply an incentive of $y\%$ if the term in the price function directly associated to the traffic congestion makes $p$ decrease of $y\%$ during the intervals of peak traffic congestion.    
\item $\chi$: this normalizing factor takes different values along the prediction horizon. Specifically, $\chi(t)=1/(W+1)$ for the first $W+1$ time intervals in $\mc T(k)$ and  $\chi(t)=1/(2W+1)$ for the others. This is due to the definition of and constraints on $\vartheta_i$ (see \cite{cenedese:2020:highway_control_pI} for more details).
\end{itemize}
\begin{table}[t]
  \centering
  \caption{The values  of the parameters in the decision making process for the base case.}
\label{tab:C0_values}
  \begin{tabular}{l  c c }
  \toprule
  \multicolumn{3}{c}{Base case: Game parameters}\\
\midrule
$x_i^{\tup{ref}}$ & -- & $[0.25,0.35]$\\
$p_{\tup{EV}}$ & -- & $5\%$ \\ 
$lT$ & $[h]$ & $0.0278$ \\
$T_{\tup h}$ & -- & $15$\\
$C_i$ & $[\tup{kWh}]$  & $\{12,\cdots,93.4,100\}$\\
$\eta_i$ & $[\tup{1/kWh}]$ &  $[0.85,0.99]$ \\
$\bar \delta$ & -- & $100$ \\
$\alpha_i$ & -- &  $(0,1)$\\
$u^{\tup{max}}$ &$[\tup{kWh}]$  & $416.6$ \\
$\overline u (\underline u)$& $[\tup{kWh}]$ & $4.16$ \\
$d$ & $[\tup{MWh}]$ & $[3.94,4.16]$ \\
$c_1,c_2$ & $[$\euro$\tup{/kWh}^2]$ & $5.07\cdot 10^{-5}$\\
$c_3$ & $[$\euro$\tup{/(h}\cdot\tup{kWh)}]$ & $0.33$\\
$\overline p_i$ & $[$\euro$\tup{/kWh}]$ &  $0.205$\\
$\beta_0$ & $[$\euro$\tup{/kWh}]$ & $0$\\
$\beta_1$ & $[$\euro$\tup{/(h}\cdot\tup{kWh)}]$ & $0.3315$ \\
$W$ & -- & $3$ \\  
$\gamma$ & $[\tup s]$ & $1.79$  \\
$\chi$ & $[\tup h]$ & $\{0.143,0.25\}$  \\
$\upsilon$ & -- & $1$ \\
\bottomrule
\end{tabular}
\end{table}
 Finally, let us suppose that the percentage of the \gls{PEV} users that are traveling with an insufficient \gls{SoC}  is very low, since we focus on drivers commuting during the rush hours so that the optimization problem is always feasible. From a theoretical  point of view, one can prevent feasibility problems by increasing the  length of the prediction horizon $T_{\tup{h}}$ or by translating the hard constraint on the minimum \gls{SoC} into a soft one. In a real scenario, this would not be an issue, since these users will simply extend their staying at the \gls{CS}.
\section{Numerical results}
\label{sec:simulation_section}
In this section, we present and discuss the simulation results for the base case introduced in the previous section (see Table~\ref{tab:C0_values}), focusing on the benefit of the \gls{ATDM} strategy in terms of the average total travel time for all the drivers.
\subsection{Performance index}
The goal of the proposed policy is to alleviate traffic congestion during the rush hours for \emph{all} the drivers on the highway, not only for the \gls{PEV} owners. To quantify the benefit (and performance) of the proposed policy, we first define the quantity $\Delta_0\coloneqq\sum_{\ell=1}^N \Delta_{0,\ell}$, which denotes the sum of the  extra travel times $\Delta_{0,\ell}$ of the drivers in the absence of the \gls{CS}. Furthermore, we define $\Delta\coloneqq \sum_{\ell=1}^N \Delta_{\ell}$ as the extra travel time in  the presence of the \gls{CS} and supposing that \glspl{PEV} subscribe the proposed policy. We then introduce   the following performance  index:
\begin{equation}
\label{eq:perf_index}
\pi(\Delta_0,\Delta) \coloneqq
\frac{\sum_{k\in\N} \Delta_0(k)-\Delta(k)}{\sum_{k\in\N}\Delta_0(k)} \cdot 100.
\end{equation}
The larger $\pi$, the greater the benefit emerging from the  behavior of \gls{PEV} owners. 
The index \eqref{eq:perf_index} is a good yardstick to evaluate the performance of the policy over the whole day and whether or not its adoption achieves the desired traffic congestion alleviation. 
\subsection{Simulation base case}
\label{sec:sim_base_case}
\begin{figure}[t]
\centering
\includegraphics[scale=0.9,width=\columnwidth]{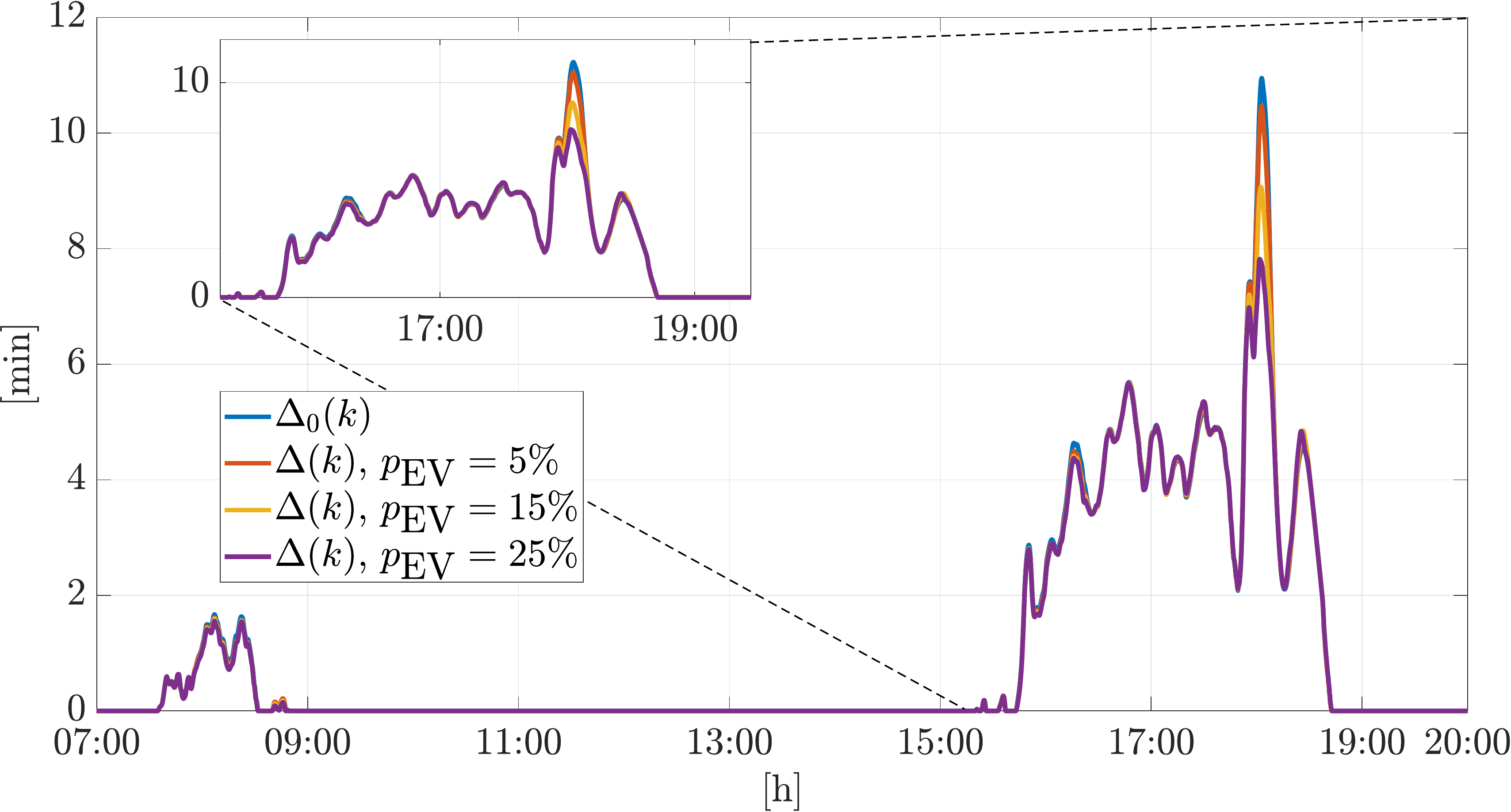}
\caption{Comparison of the extra travel time  due to the congestion in the  presence ($\Delta$) and absence ($\Delta_0$) of the \gls{CS}, respectively. The simulations are carried out for different values of $p_{\tup{EV}}$.}
\label{fig:Delta}
\end{figure}

We are now ready to discuss the effects of the \gls{ATDM} strategy proposed in (Part I: Theory) \cite{cenedese:2020:highway_control_pI}. 
Figure~\ref{fig:Delta} highlights that there are  two time intervals during  which a traffic congestion occurs and they are centered around $08$:$00$ and $18$:$00$, respectively, and reflect the real data in Figure~\ref{subfig:vel}, where the velocity drops exactly in those time intervals. The traffic jam in the morning lasts fro  almost one hour while the one in the afternoon for three hours, with a particularly high peak (in terms of travel time) at $18$:$00$.
Remarkably, despite the percentage of \glspl{PEV} on the highway is modest, our algorithm manages to decrease the afternoon peak of more than $5.4\%$ in terms of $\Delta$. If there is no congestion all the cells are traveled by a vehicle in $1.62\,\tup{min}$. During peak traffic congestion this time becomes $12.54\,\tup{min}$, the proposed \gls{ATDM} strategy reduces it to $12.06\,\tup{min}$.  
The overall achievement in terms of travel time is  $ \pi =1.05$. 
\begin{figure}[t]
\centering
    \subfloat{{\includegraphics[scale=0.9,width=0.49\columnwidth]{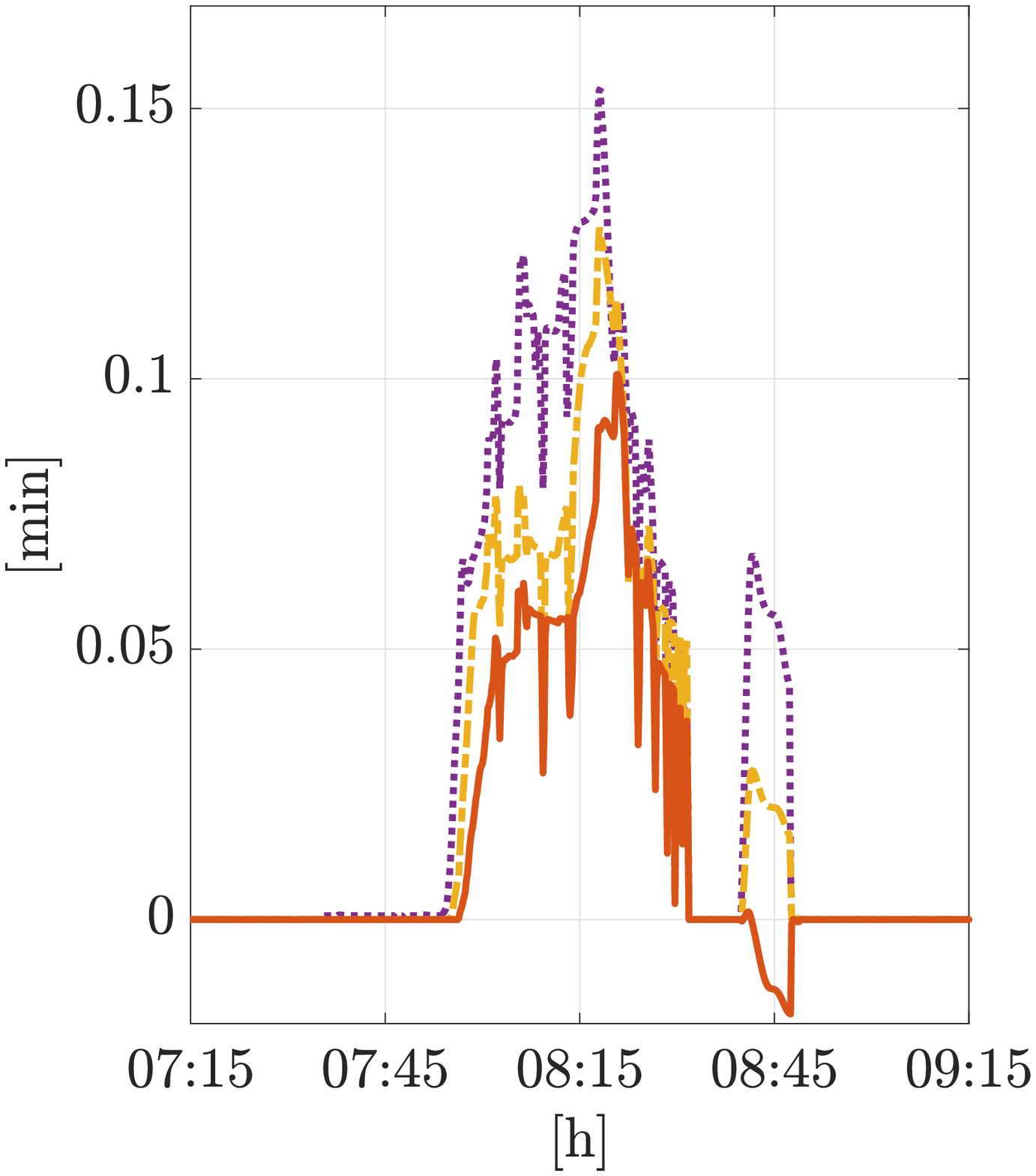}}}%
    \subfloat{{\includegraphics[scale=0.9,width=0.455\columnwidth]{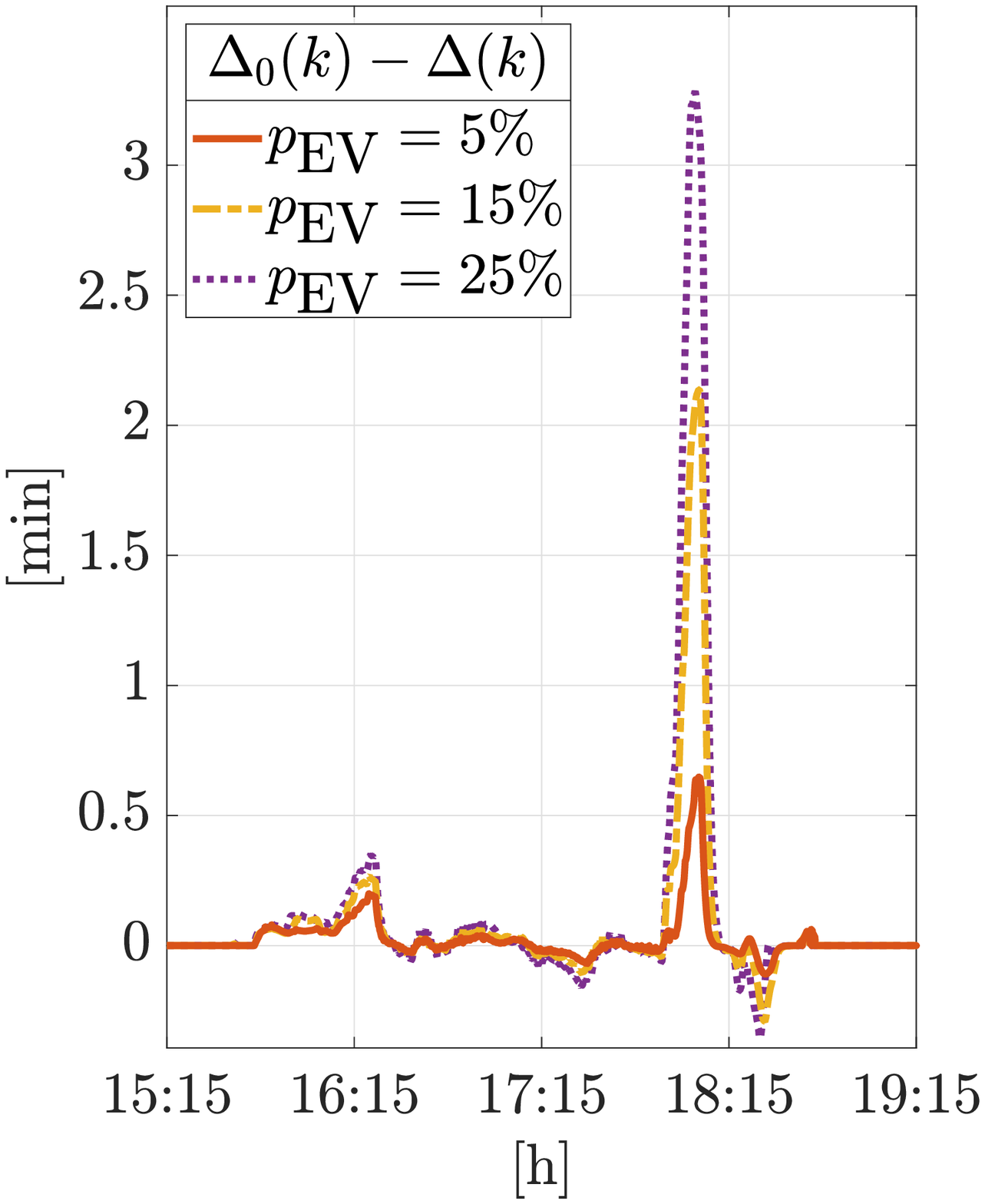} }}%
\caption{Comparison of the difference $\Delta_0-\Delta$ computed with different values of $p_{\tup{EV}}$. The two figures show the detail related to the morning and afternoon congestion, respectively.}
\label{fig:diff_Delta}
\end{figure}
In Figure~\ref{fig:diff_Delta}, we highlight the difference between the travel time with vs without our \gls{ATDM} strategy. The \gls{ATDM} manages to decrease the peaks and redistribute the vehicles in the ``valleys'', i.e., the parts in which the curve  takes lower values with  respect to the closest peaks. This emerging behavior matches the expectation: in fact, it relates to the ``valley filling'' phenomenon, well known in the electricity market, where the energy price depends on the energy demand, see \cite{J_Hiskens_2013,cenedese:2019:PEV_MIG}. The redistribution of the vehicles creates a global reduction of the travel time. This effect of the \gls{ATDM} on the \glspl{PEV} is confirmed by the flows shown in Figure~\ref{fig:r2s_s2r}, where  \gls{r2s}  increases  before the two peaks of $\Delta$, while \gls{s2r} peaks immediately after the reduction of  $\Delta$ after the rush hours.
\begin{figure}[t]
\centering
    \subfloat{{\includegraphics[scale=0.9,width=\columnwidth]{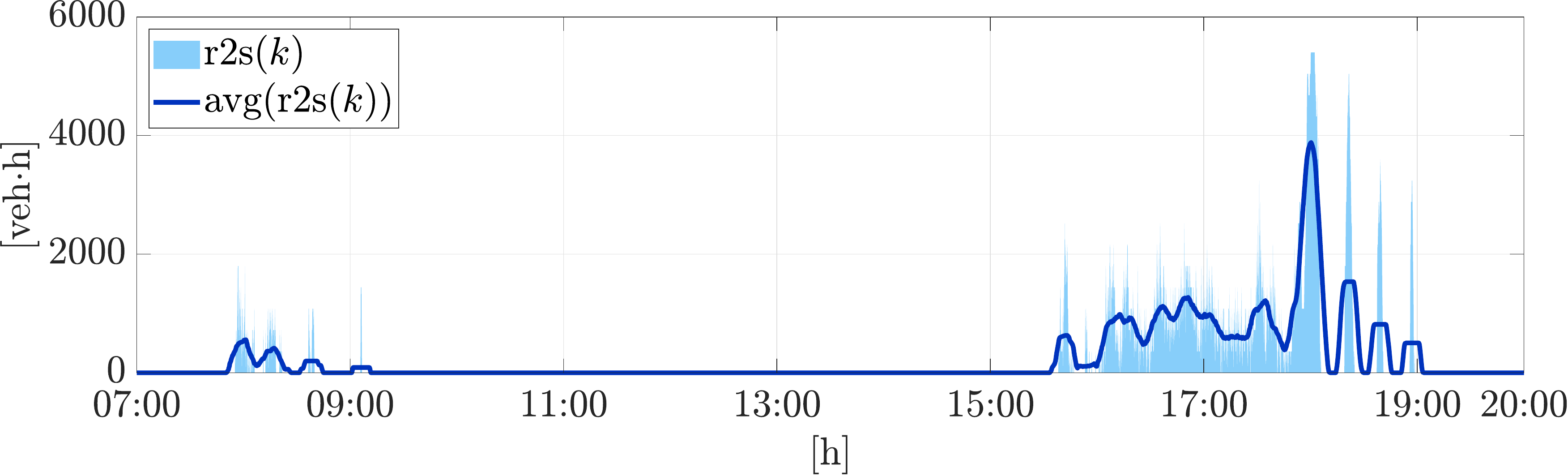}}}\\%
    \subfloat{{\includegraphics[scale=0.9,width=\columnwidth]{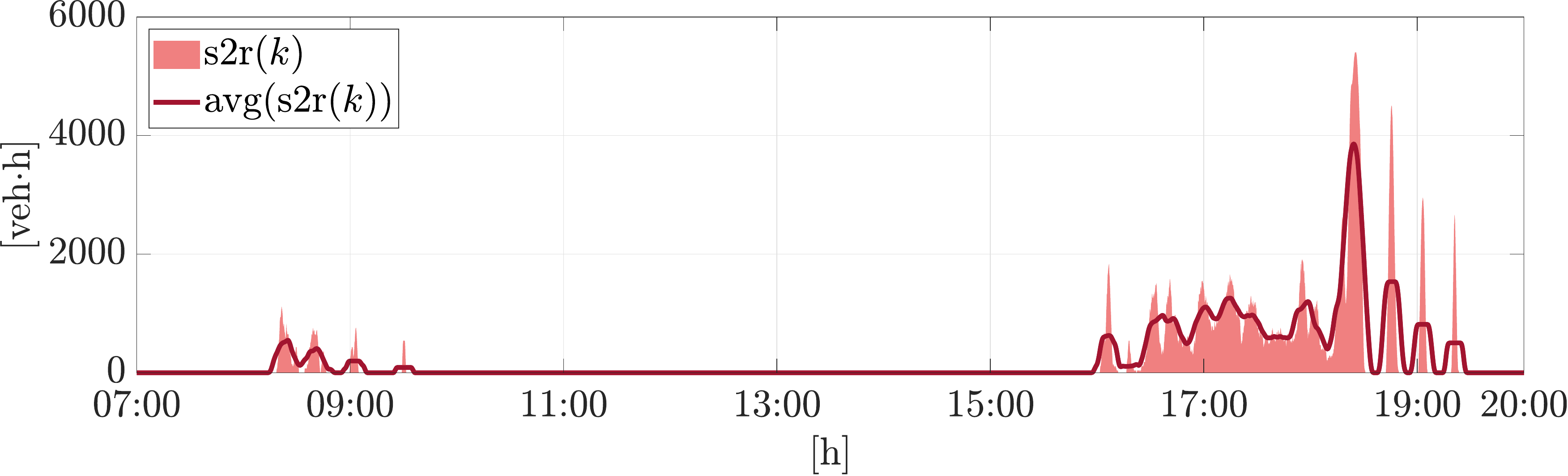} }}%
 \caption{The flows entering and exiting the \gls{CS}, i.e., \gls{r2s} and \gls{s2r} respectively. The shaded area represents the actual flow, while the solid line a moving average  with a window of $10\,\tup{min}$.}
\label{fig:r2s_s2r}
\end{figure}

\begin{figure}[t]
\centering
\includegraphics[scale=0.9,width=\columnwidth]{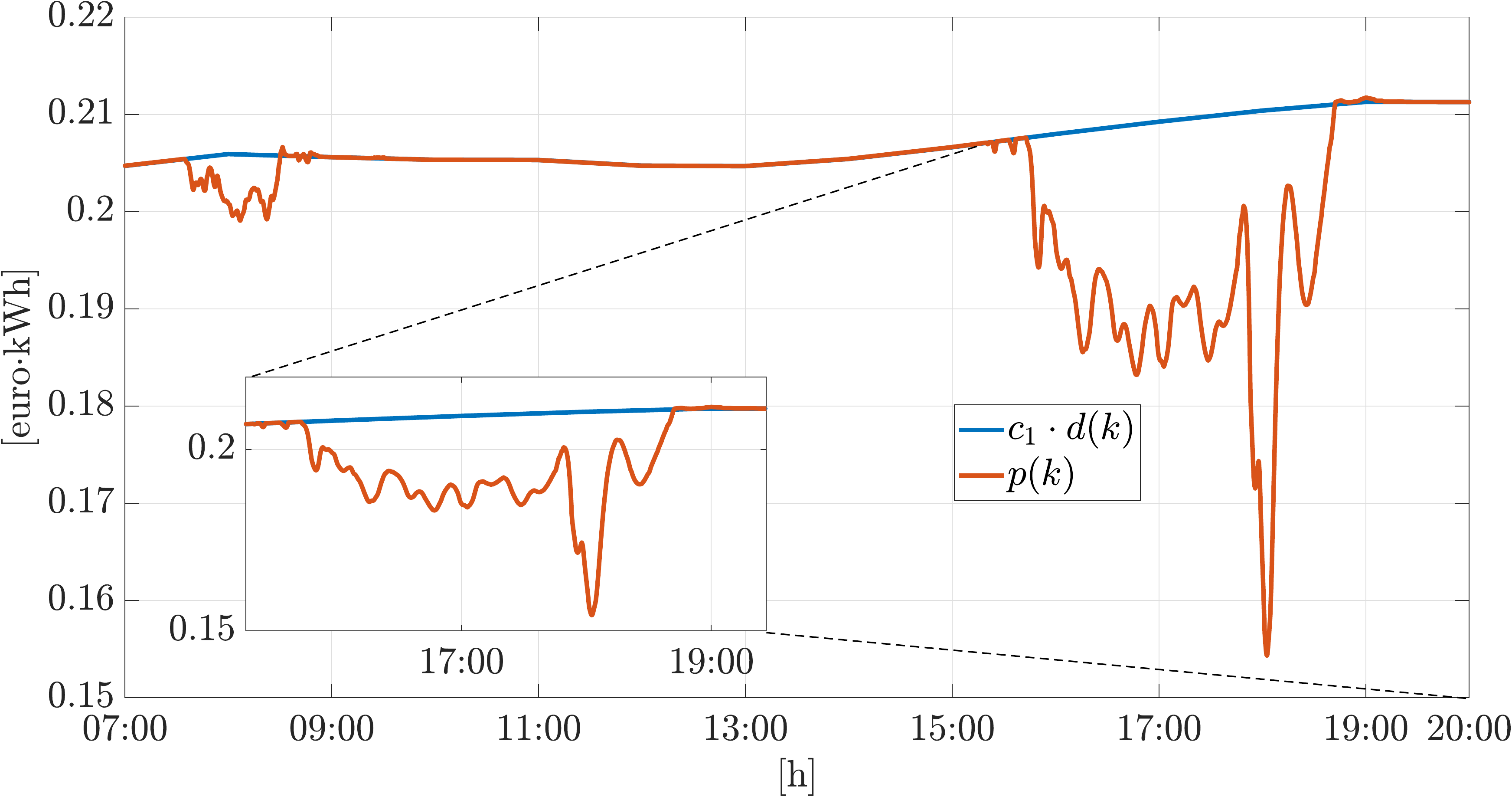}
\caption{The energy price  $p$ adopted by the \gls{CS} and the base demand $d$ scaled by the factor $c_1$. }
\label{fig:price}
\end{figure}

The dynamic energy price $p$ designed by the \gls{HO} is reported in Figure~\ref{fig:price}. It shows a slow variation due to the base energy demand $d$ and a fast one associated with the traffic congestion on the highway stretch. 
Interestingly, these two quantities act in opposite directions since the major traffic congestions happen during the rush hours, which are notoriously  those during which the demand of energy is at its highest. The parameters $c_1,c_2$ and $c_3$  are chosen such that the incentive provided is the $25\%$, i.e., the maximum discount is equal to $0.05 \, $\euro$\tup{/kWh}$ with respect to  the average price  $\bar p_i=0.205\, $\euro$\tup{/kWh}$. 
\begin{figure}[t]
\centering
\includegraphics[scale=0.9,width=\columnwidth]{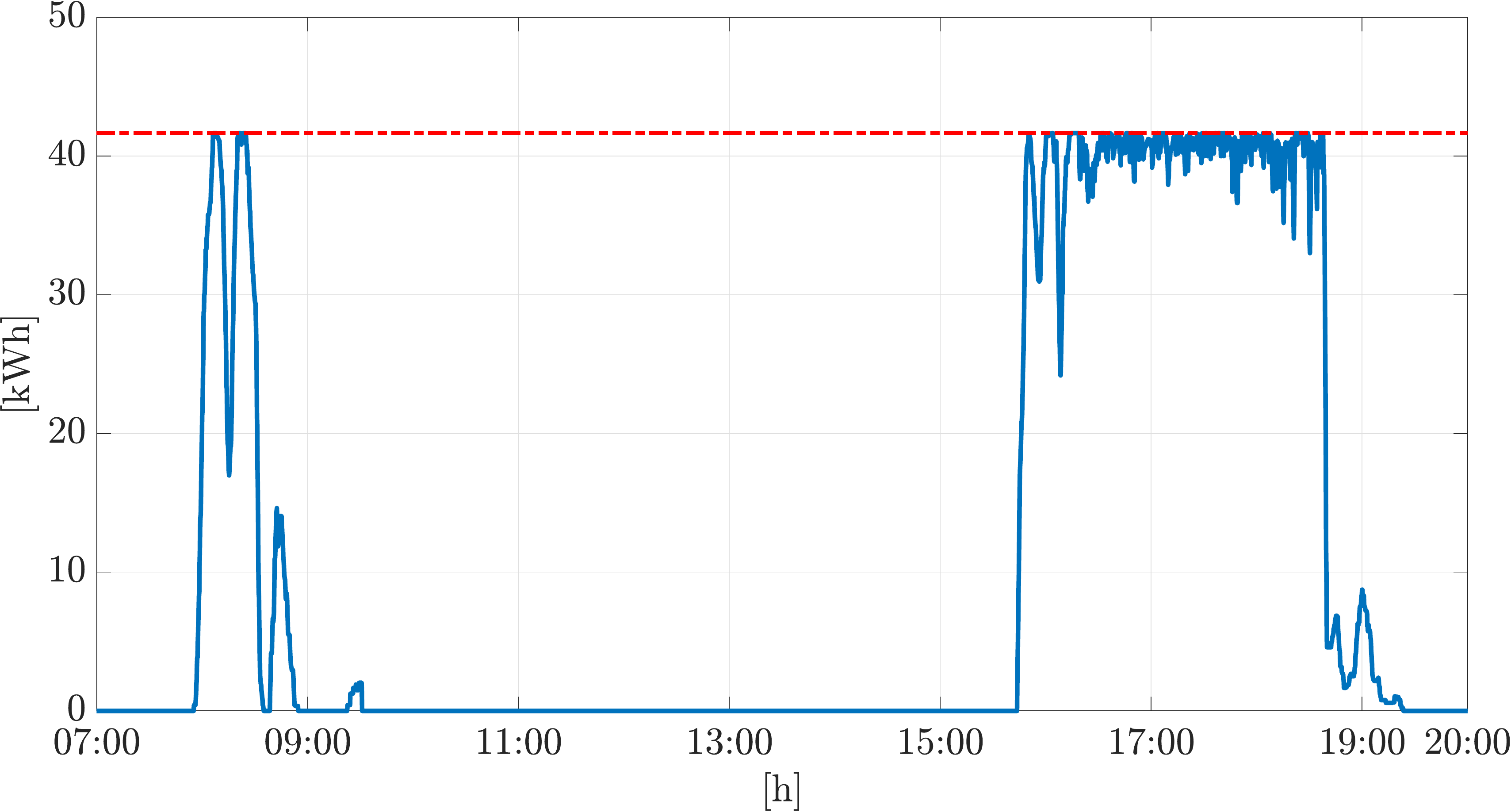}
\caption{The total energy purchased by the \glspl{PEV},  $u^{\tup{PEV}}$. The red dashed line denotes $u^{\tup{max}}$.}
\label{fig:u_PEV}
\end{figure}
\begin{figure}[t]
\centering
\includegraphics[scale=0.9,width=\columnwidth]{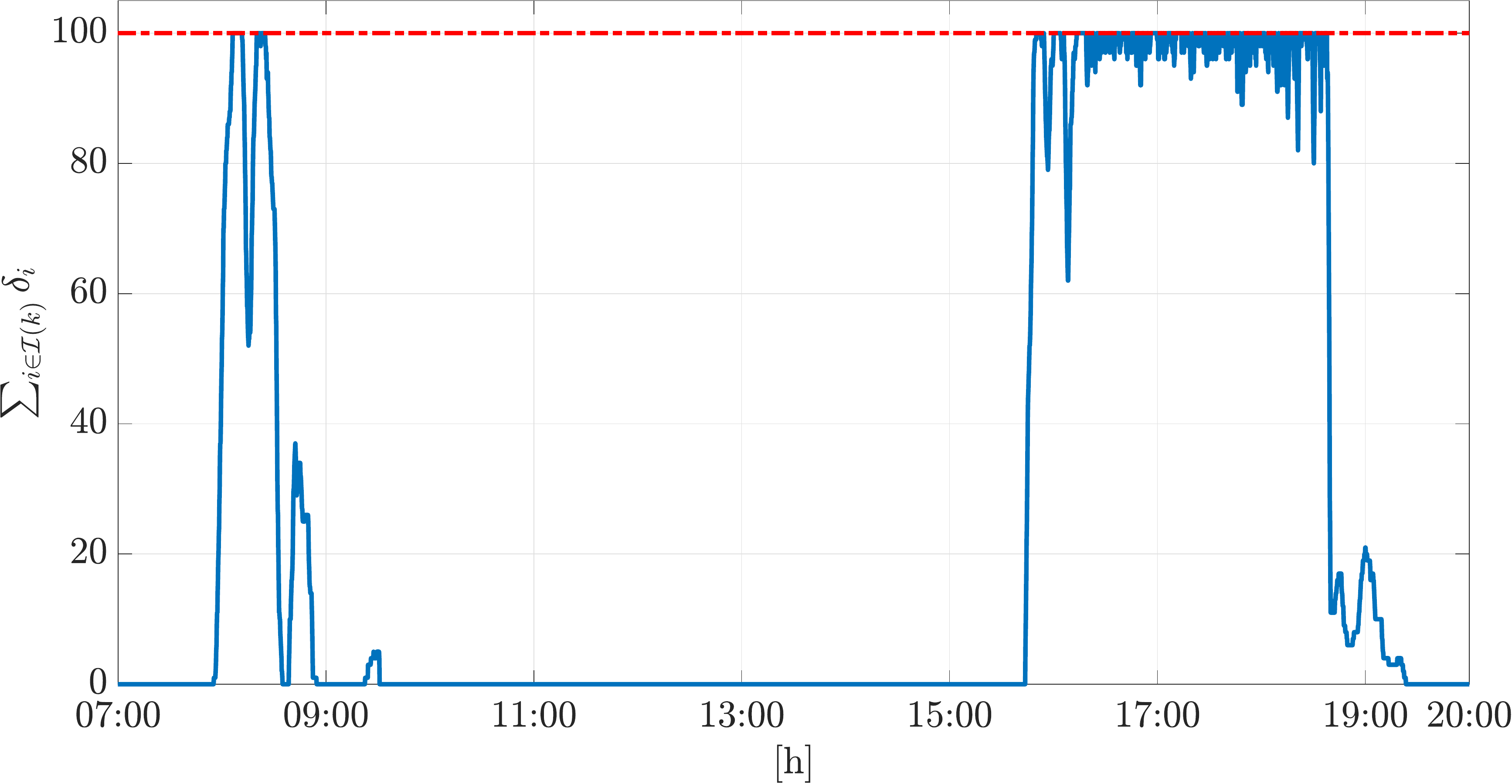}
\caption{The number of busy charging spots at the \gls{CS}. The red dashed line is set equal to $\bar \delta=100$, the capacity of the \gls{CS}.}
\label{fig:delta}
\end{figure}
Finally, in Figures~\ref{fig:u_PEV} and \ref{fig:delta}, we illustrate the number of \glspl{PEV} simultaneously connected to the grid, i.e.,  $\sum_{\bar k<k}\sum_{i\in\mc I(\bar k)} \delta_i(\bar k)$, and the amount of energy purchased $u^{\tup{PEV}}$. These two quantities are closely related and for this reason their profiles are almost identical. In both cases, the strategies implemented satisfy the upper bound constraints. As expected, during the period of high traffic congestion all the charging spots at the \gls{CS} are busy. This  highlights a key difference between longer and shorter traffic congestion; in the latter, there are only brief periods during which no charging spot is available, i.e., $\sum_{\bar k<k}\sum_{i\in\mc I(\bar k)} \delta_i(\bar k)=\bar \delta$. Therefore, if a \gls{PEV} decides to enter the \gls{CS}, then it is almost sure to connect  immediately. On the other hand, a traffic congestion that lasts for a substantial amount of time creates a queue of \glspl{PEV} at the \gls{CS}  waiting for charging. This idle waiting time creates an extra cost for the \glspl{PEV} stopping.  Moreover, this phenomenon generates  detrimental effects in the concurrence of high traffic congestion peaks surrounded by an overall elevated traffic congestion level. In fact, if we consider \glspl{PEV} exiting cell $1$ at $18$:$10$, then it should be beneficial for them to stop and charge for a short period, waiting for the traffic congestion to disappear. Unfortunately, all the charging spots are occupied, so the \gls{PEV} cannot stop, hence contributing   to increase the traffic congestion. In fact, Figure~\ref{fig:r2s_s2r} shows that in concurrence with the highest congestion peak, the \gls{r2s} flow decreases. 

This first example highlighted the overall benefit of adopting a dynamic energy price to enforce an \gls{ATDM}. The most important macroscopic effect is the redistribution of the vehicles that creates a valley filling like phenomenon in the travel time.     
\section{Sensitivity analysis}
\label{sec:sensitivity}
Next, we investigate the effects that a  single parameter has on $\Delta$. They are difficult to extrapolate directly from the base case due to the complexity of the decision making process and the multitude of variables involved. In this section, we propose a sensitivity analysis that focuses on the most compelling features of the proposed \gls{ATDM}. Specifically,  we   start from the parameter configuration in Table~\ref{tab:C0_values} and vary one parameter (or at most two) to highlight how it affects the travel time. Note that the values used for this variation may deviate from what one can reasonably expect in a real-world scenario, e.g. a \gls{CS} with $\bar \delta=1000$, but they are meant to investigate the performance in extreme configurations or approximations of possible future configurations, e.g. one \gls{CS} that approximates the charging capacity of two or more nearby \glspl{CS}.
\subsection{Number of available charging spots}
\label{sec:ch_spots}
Intuitively, a larger number of charging spots $\bar \delta$ mitigates the effects of  prolonged traffic congestion; in fact, the \glspl{PEV} take more time to saturate the available spots at the \gls{CS}, and, as a consequence, the congestion alleviation during the rush hours can be more effective. 
\begin{figure}[t]
\centering
\includegraphics[scale=0.9,width=\columnwidth]{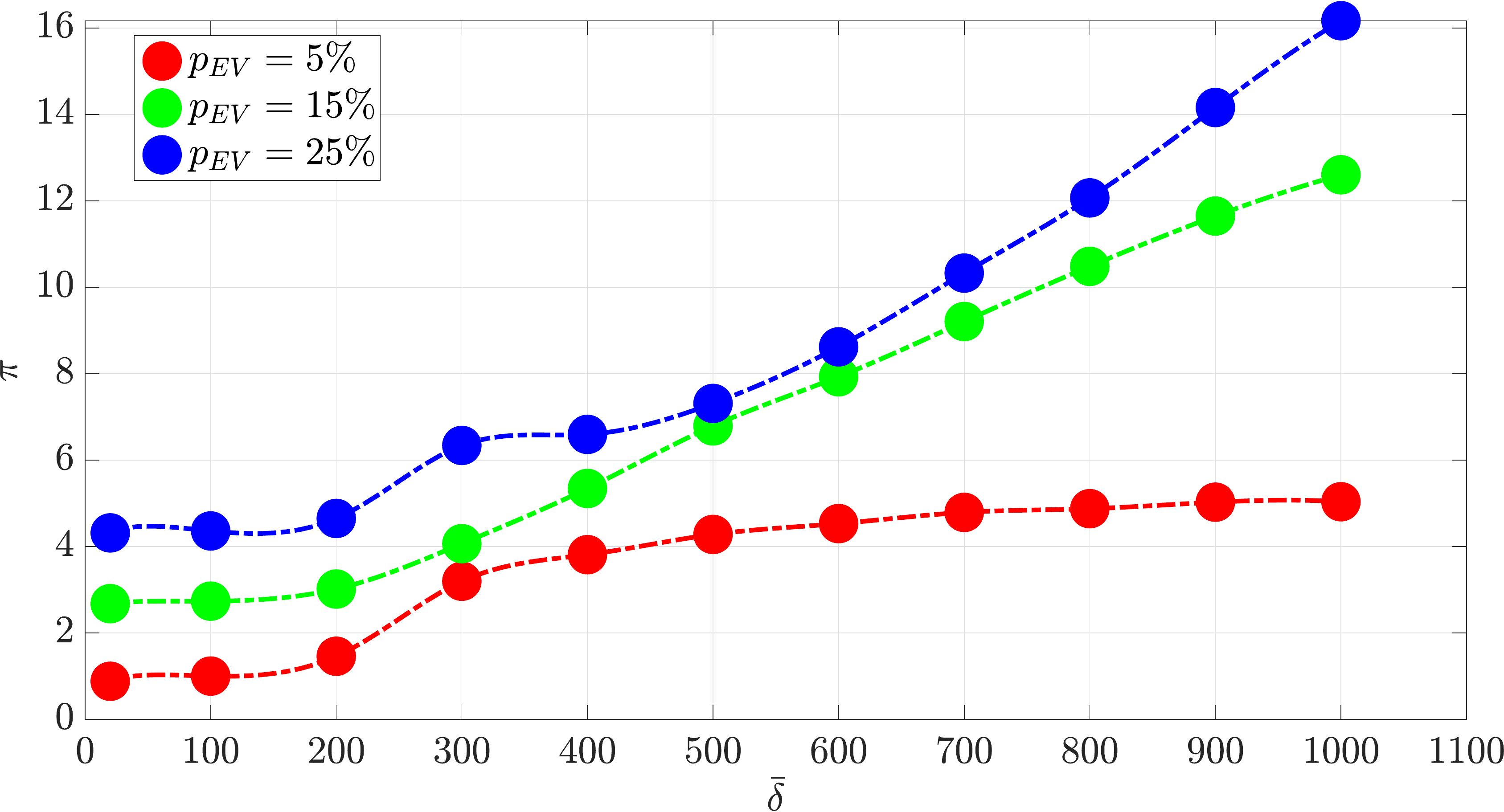}
\caption{Effect of the increment $\bar \delta$ on the performance index $\pi$ in \eqref{eq:perf_index}. The simulations are carried out for different values of $p_{\tup{EV}}$.}
\label{fig:plugs_vs_p_ev}
\end{figure}
The variation of $\pi$ with respect to the increment of charging spots is presented in Figure~\ref{fig:plugs_vs_p_ev} for different \gls{PEV} shares, i.e.,  $p_{\tup{EV}}$. The curve profile for $p_{\tup{EV}}=5\%$ shows two plateaus, one for small values of $\bar\delta$ and the other for values greater  than $400$. The transit between the two happens between $\bar\delta=200$  and $\bar\delta=300$. It represents the setting in which an investment to increase the number of charging spots is the most profitable in terms of performance.
 In fact, going from $\bar\delta= 100$  to $\bar\delta=300$ triples the value of $\pi$, which rises from $1.05$ to $3.19$.
 
Assuming a higher number of \glspl{PEV} on the highway does not avoid the initial plateau, but, surprisingly, it prevents the occurrence of an analogous plateau for greater values of $\bar\delta$. This implies that investing in more charging spots creates a high return in terms of performance, after a certain ``critical'' value, i.e., the value of $\bar{\delta}$ after which the initial plateau ends.  Interestingly, even if  $p_{\tup{EV}}$ varies, the critical value of $\bar\delta$ remains between $200$ and $300$. In Figures~\ref{fig:Delta} and \ref{fig:diff_Delta}, we note that the beneficial effect of an increment of the \glspl{PEV} number. In fact, the more \glspl{PEV} subscribe the higher the traffic alleviation is. Figure~\ref{fig:diff_Delta}, clearly shows the valley filling effect, namely a higher peak reduction implies a subsequent increment of travel time during the less-congested period.   
\subsection{Charging spots and incentive}
\label{sec:plugs_price}
Starting from the parameters in Table~\ref{tab:C0_values}, we analyze the performance obtained by varying simultaneously the number of available charging spots, i.e., $\bar \delta$, and the incentive on the energy price. 
\begin{figure}[t]
\centering
\includegraphics[scale=0.95,width=\columnwidth]{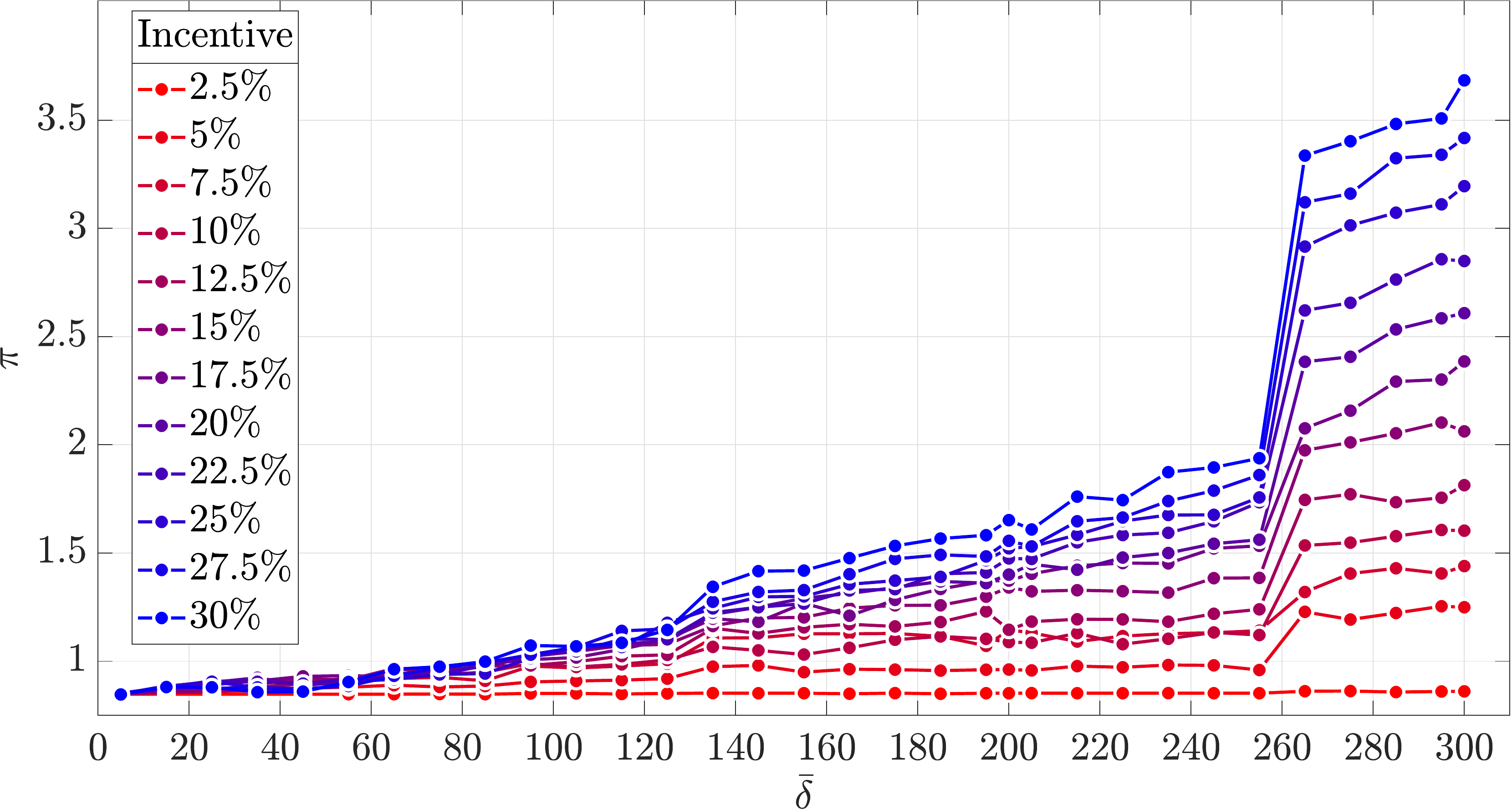}
\caption{The performance $\pi$ as a function of  the number of charging spots, i.e., $\bar \delta$, computed for different incentives.}
\label{fig:plugs_price}
\end{figure}
Figure~\ref{fig:plugs_price} depicts the different curves resulting by increasing the number of available charging spots, i.e., $\bar\delta\in[5,295]$, when the incentive ranges between  $2.5\%$ and $30\%$. An increment in the number of plugs generates, in general, an higher performance increment compared to a greater incentive. Surprisingly, the value of $\bar\delta$ at which the initial plateau of performance stops is not affected by the incentive. In fact, the consequent  steep increment in $\pi$ settles between $\bar\delta=255$ and $265$; this can be considered the critical value for this configuration. On the other hand, an increment in the incentive creates a smaller effect  on the performance yet it is more predictable  especially if the number of plugs is high. This result suggests that an investment in increasing $\bar\delta$ leads to a high return only if the additional charging spots allow to surpass the critical level. If this is not possible, then increasing the incentive is a more profitable strategy.
\subsection{Heterogeneous population of \glspl{PEV}}
\label{sec:hetero_pop}
In the simulations performed for the base case in Section~\ref{sec:sim_base_case}, we have considered a fairly homogeneous population of \gls{PEV} owners, i.e., drivers that value similarly their interest of saving time or money. The probability distribution, from which all the $\alpha_i$ are drawn, has variance $\sigma_\alpha =0.03$. Next, we consider a collection of heterogeneous \glspl{PEV}, in which some are more interested in maintaining the travel time at its minimum and others in saving money. 
\begin{figure}[t]
\centering
\includegraphics[scale=0.9,width=\columnwidth]{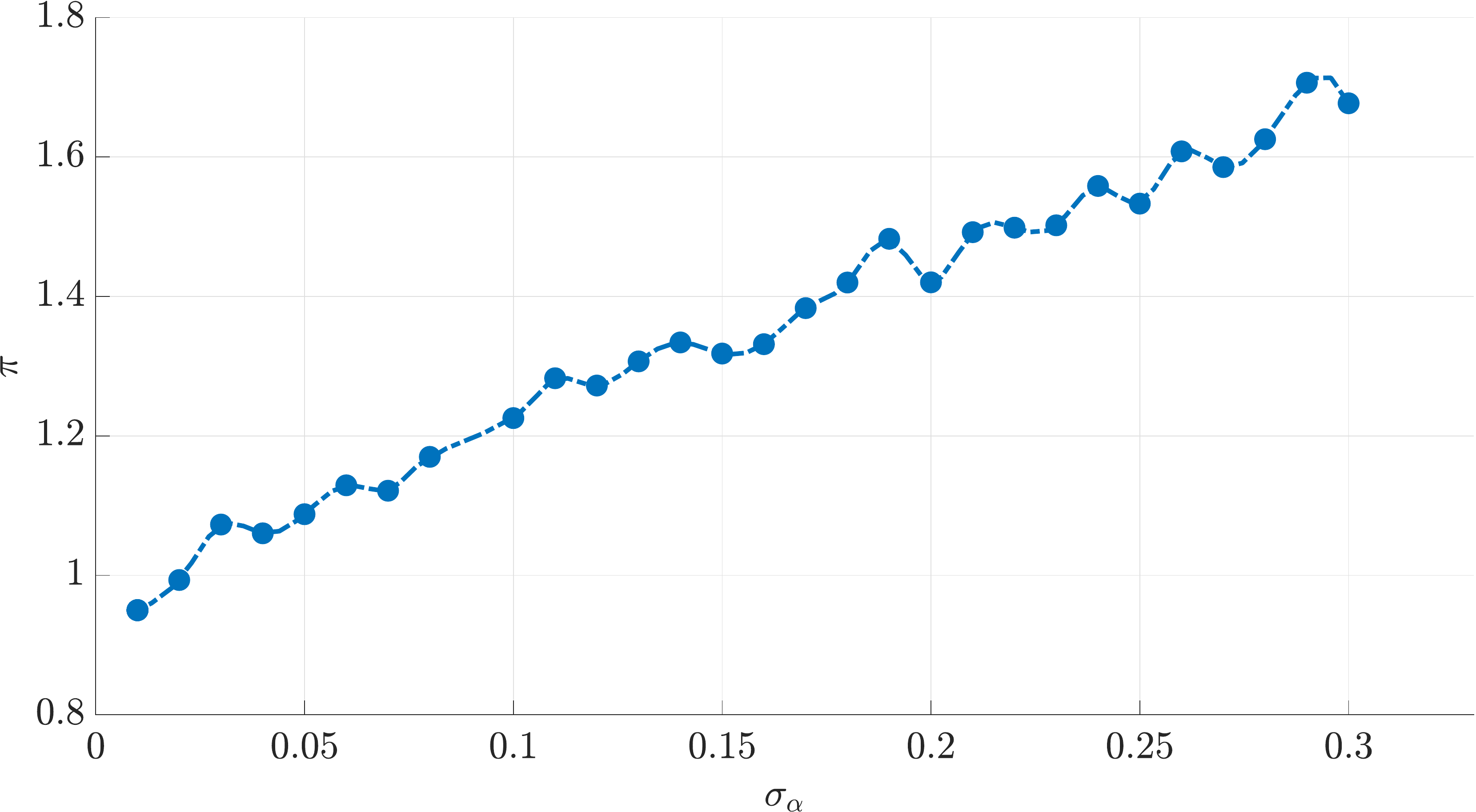}
\caption{Performance $\pi$ as a function of  $\sigma_\alpha$, the heterogeneity in drivers interests.}
\label{fig:alpha_var}
\end{figure}
In Figure~\ref{fig:alpha_var}, we show the performance increment attained by assuming a population of agents with increasingly diverse interests (all the other parameters are set as in Table~\ref{tab:C0_values}). Interestingly, the value of $\pi$ increases from $0.95$, when $\sigma_\alpha=0.01$, all the way to $1.71$, if  $\sigma_\alpha=0.29$. The trend of the increment is almost linear and does not seem to slow down for higher values of $\sigma_\alpha$. From \cite{verhoef:2010:incentive_traffic_managment}, it seems reasonable to assume that the commuters involved in the decision making process form a heterogeneous set; nevertheless, the fact that we focus only on \gls{PEV} owners  may reduce this variability. The value of $\alpha$ that describes the best a real-world scenario is for these reasons an open problem that requires further investigation.  
The increment in the performance can be explained by looking at Figure~\ref{fig:delta}. In fact, a more heterogeneous population leads to fewer \glspl{PEV} stopping at the \gls{CS} during the long afternoon congestion peak. This alleviates the problem of overcrowding the \gls{CS} where all the charging spots are generally busy during the largest traffic congestion peaks.
\subsection{Incentive}
\label{sec:incentive}
Based on our numerical study, we have chosen the incentive used in the base case heuristically. In this subsection, we study the effect of different incentives if $p_{\tup{EV}}$ varies. 
\begin{figure}[t]
\centering
\includegraphics[scale=0.9,trim = {80 0 50 30 },clip,width=\columnwidth]{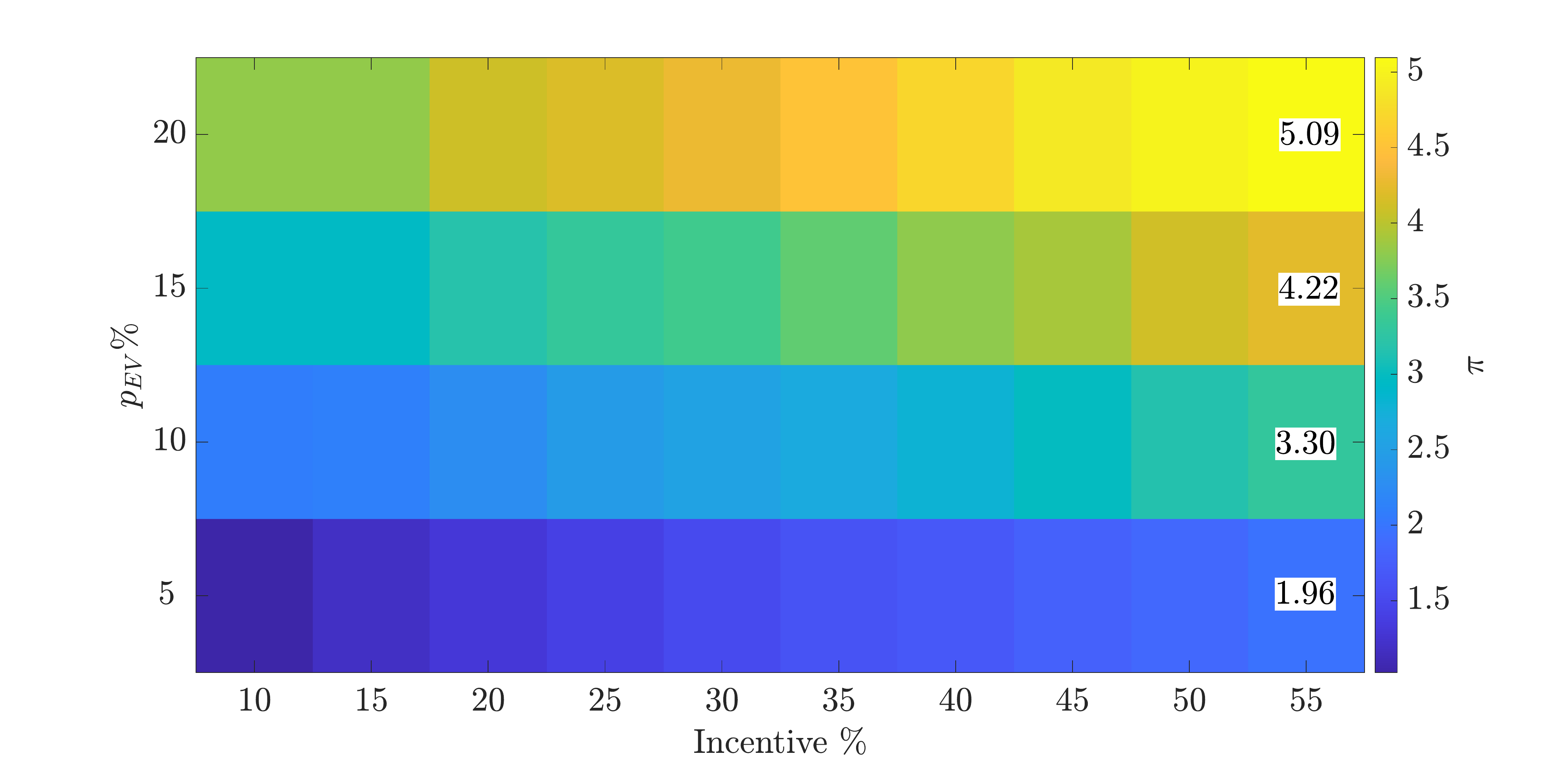}
\caption{The performance $\pi$ in \eqref{eq:perf_index} for different combinations of $p_{\tup{EV}}$ and percentage of  incentive. We explicitly reported the value of $\pi$ only for the best performer of each row.}
\label{fig:price_pev}
\end{figure}
In Figure~\ref{fig:price_pev}, we report the performance $\pi$  computed for all the combinations of the incentive $[10,55]$ and  $p_{\tup{EV}}\in\{5,10,15,20\}$. If $p_{\tup{EV}}=5\%$ the value of $\pi$ is almost constant with fluctuations of $\pm 0.1$. In the other cases, an interesting trend arises: the optimal incentive that maximizes the performance is not the highest one. In fact, if $p_{\tup{EV}}=10\%$ the optimal incentive is $15\%$ while for $p_{\tup{EV}}=15\%$ and $p_{\tup{EV}}=20\%$ it is $35\%$. This relates to the phenomena described at the end of Section~\ref{sec:sim_base_case}. Inflating the incentive makes the number of \glspl{PEV} willing to stop grow. This has a beneficial effect on the morning peak, but a detrimental one on the afternoon rush hour, since it exacerbates the problem of the overly crowded \gls{CS} and the lack of available charging spots. 
\begin{figure}[t]
\centering
\includegraphics[scale=0.9,trim = {80 0 50 30 },clip,width=\columnwidth]{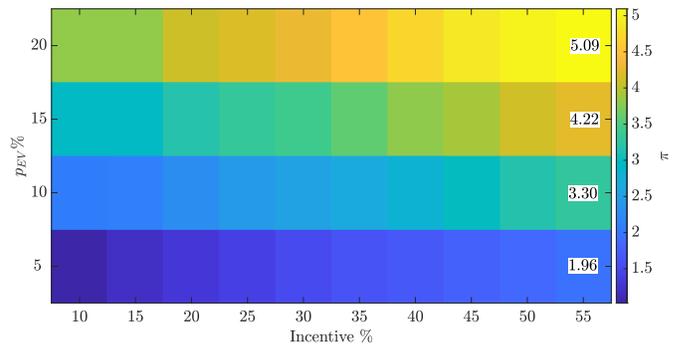}
\caption{The performance $\pi$  in \eqref{eq:perf_index} for different combinations of $p_{\tup{EV}}$ and percentage of  incentive. The incentive takes the nominal value during the morning peak and $1/5$ of it during the afternoon peak. We explicitly reported the value of $\pi$ only for the best performer of each row.}
\label{fig:dyn_price_pev}
\end{figure}
To mitigate this issue, we propose in the next simulation a discount that varies over time and, in particular,  takes a higher value during the morning traffic congestion peak and a lower value during the afternoon one. Specifically, if  we say we apply an incentive of  $y\%$, then it means that we apply a $y\%$ incentive from $7$:$00$ to $16$:$00$, and $0.2\, y\%$ for the remainder of the day. Implementing this smarter type of incentive leads to striking results. In Figure~\ref{fig:dyn_price_pev}, for $p_{\tup{EV}}=5\%$ the travel time significantly decreases with the increment of the incentive, i.e., $\pi$ goes from $1.27$ (see Figure~\ref{fig:price_pev}) to $1.96$. This trend holds also for higher values of  $p_{\tup{EV}}$, increasing the overall maximum performance $\pi$ from $4.3$ to $5.09$. This suggests that the problem of over incentivizing the \glspl{PEV}  to stop during the afternoon should be solvable with incentives decreased for the afternoon peak only. Finally, the fact that the maximum value  of $\pi$ always coincides with the maximum incentive suggests that there  is still room for improving the performance by increasing the incentive even further.
Nevertheless, one should be careful in implementing a very large incentive, because it might be overly costly for the \gls{HO} that has to take care of the economical burden associated with the energy price discount.
\section{Policy recommendation}
\label{sec:policies}
Let us now extrapolate the key messages  based on the numerical analysis and results of the previous sections, and then  elaborate on possible policies that can  improve the performance of the \gls{ATDM}.

As illustrated in the previous section, tuning the parameters involved in the decision making process translates  into  \textit{infrastructural} or \textit{sociological} actions. The former are related to tangible features of the infrastructure, e.g., the number of available charging spots $\bar \delta$ or the maximum power that fast charging can deliver $\bar u$; the latter instead are intrinsically linked to the behaviors and decisions of the drivers, e.g., the interests of each agent $\alpha_i$ or the incentive applied to steer the drivers' behavior. As expected, the infrastructural improvements create greater outcomes in terms of performance, as shown in  Figure~\ref{fig:plugs_price} where the performance improvement by increasing the number of available charging spots overcomes that  due to a higher monetary discount. This type of interventions is expensive and produces inelastic changes with respect to  the demand. As an illustrative example, let us focus on $\bar\delta$:  increasing  it leads to a \gls{CS} that has a low percentage of charging spots occupied throughout most of the day, and  thus the return of the investment is limited. This economic point of view and the existence of a critical value of $\bar\delta$ (see Section~\ref{sec:ch_spots}) suggest that the \gls{HO} should carefully decide whether it is convenient to invest in this direction. Infrastructural actions should be rare and performed when a substantial leap forward can be achieved, while small incremental improvement should be avoided. 

Enforcing a change in the drivers' behavior is not as straightforward as changing some physical features. Nevertheless, our formal analysis and modeling of the decision making process help us to infer some possible policies to achieve an improvement in the traffic congestion alleviation. The energy price discount generates remarkable performance, if it is designed by the \gls{HO} based on the severity of the  traffic congestions, see Figures~\ref{fig:price_pev}~and~\ref{fig:dyn_price_pev}. This policy does not require massive investment to be implemented and it is elastic with respect to the current traffic situation. Thus, from an economic perspective, a dynamic energy price currently seems the safest and the  most remunerative option for the \gls{HO}. The growing trend of incentivizing smart working and flexible working hours has a beneficial effect on the  proposed policy for \gls{ATDM}. In fact, it will increase the variety of interests among the \gls{PEV} owners and  consequently the performance should also increase, see Figure~\ref{fig:alpha_var}.  Moreover, flexible schedules may shift the overall interest of the drivers towards saving money rather than time. This variation will allow the \gls{HO} to attain the same performance even with  lower  incentives.

To summarize, the theoretical framework developed and the numerical simulations carried out in this work have convinced us that a  solid  policy to maximize the efficacy of the \gls{ATDM} over time should be composed of \textit{short term} and \textit{long term} actions  following two broad  guidelines:
\begin{itemize}
\item \textit{Short term:} a dynamic energy price should be implemented where the incentive varies with respect to  the intensity and (estimated) duration of the traffic congestion.
\item \textit{Long term:} consider a major infrastructural investment to increase the \gls{CS} capabilities, e.g., increase the number of charging spots, and consequently achieve a performance boost. In the A$13$ case study analyzed in this paper, the long term actions should be introduced, in addition to the short term ones,  when the \gls{HO} can increase the charging plugs above $100$. This ensures a substantial improvement in performance that motivates the monetary investment. 
\end{itemize}
\section{Conclusion and outlook}
\label{sec:conclusion}
The \gls{ATDM} strategy proposed  in (Part I: Theory) \cite{cenedese:2020:highway_control_pI} has been validated by simulation in a highly congested highway stretch, achieving a reduction of the peak traffic congestion of about  $5.4\%$ in terms of travel time. This result is achieved with the current value of \gls{PEV} market shares that is around the $5\%$. This method creates a redistribution of the vehicles throughout the day, achieving a travel time that is on average $1\%$ lower for all the vehicles (not only \glspl{PEV}) traveling on the highway. The foreseeable increment of the \glspl{PEV} and the strengthening of the charging infrastructure, together with the increasing popularity of flexible working schedules, will enhance the benefits of applying the proposed policy, making it even more effective over the upcoming years. 
This paper is a self-contained analysis that discovers interesting results. Nevertheless, it can act as a stepping stone to develop further research based on a similar concept. Perhaps, the most valuable  extension can be a field study in which this approach is validated with humans in the loop.  The framework of potential games well suits the introduction of partially rational agents (see for instance \cite{zino:2017:imitation,cenedese:2019:RBR})  that may introduce new algorithmic dynamics in the framework. In the literature, other \gls{ATDM} strategies have been proposed, e.g.  \cite{metropia:2019:bart2}. It can be of great value to carry out a comprehensive study that analyzes if and how well they work in combination to improve the traffic congestion alleviation.
\bibliographystyle{IEEEtran}
\bibliography{19_CDC_PEV_MC,library_CC}

%
\vspace{-1.75cm}
\begin{IEEEbiography}[{\includegraphics[trim=65 0 65 0,width=1in,height=1.25in,clip,keepaspectratio]{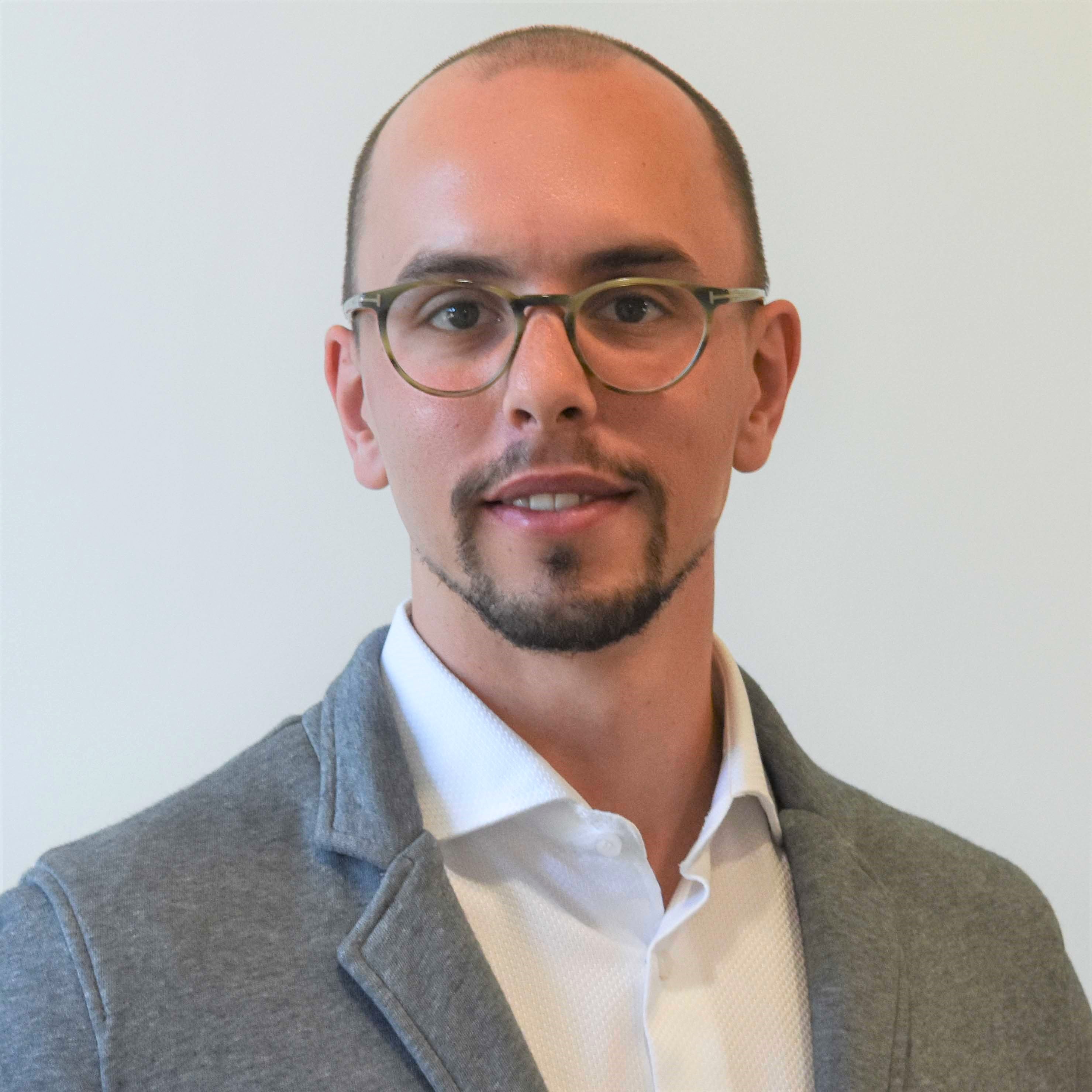}}]{Carlo Cenedese}
received the M.Sc. degree in Automation Engineering from the University of Padova, Padova, Italy, in 2016. 
From August to December 2016, he worked for VI-grade srl in collaboration with the  Automation Engineering group of the University of  Padova as a research fellow. In 2021, he completed his Ph.D. degree with the Discrete Technology and Production Automation (DTPA) Group in the Engineering and Technology Institute (ENTEG) at the University of Groningen, the Netherlands. During December 2019 he visited the Department of Mathematical Sciences, Politecnico di Torino, Torino, Italy. 
Since 2021 he is a Postdoc in the Department of Information Technology and Electrical Engineering at  the ETH Z\"urich.  
His research interests include game theory, traffic control, complex networks and multi-agent network systems associated to decision-making processes.
\end{IEEEbiography}

\vspace{-1.5cm}
\begin{IEEEbiography}[{\includegraphics[trim=0 0 0 0,width=1in,height=1.25in,clip,keepaspectratio]{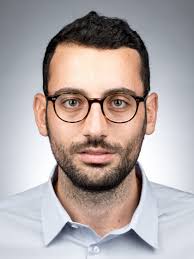}}]{Michele Cucuzzella}
received the M.Sc. degree (Hons.) in Electrical Engineering and the Ph.D. degree in Systems and Control from the University of Pavia, Pavia, Italy, in 2014 and 2018, respectively. Since 2021 he is Assistant Professor of automatic control at the University of Pavia. From 2017 until 2020, he was a Postdoc at the University of Groningen, the Netherlands. From April to June 2016, and from February to March 2017 he was with the Bernoulli Institute of the University of Groningen. His research activities are mainly in the area of nonlinear control with application to the energy domain and smart systems. He has co-authored the book Advanced and Optimization Based Sliding Mode Control: Theory and Applications, SIAM, 2019. He serves as Associate Editor for the European Control Conference since 2018 and received the Certificate of Outstanding Service as Reviewer of the IEEE Control Systems Letters 2019. He also received the 2020 IEEE Transactions on Control Systems Technology Outstanding Paper Award, the IEEE Italy Section Award for the best Ph.D. thesis on new technological challenges in energy and industry, the SIDRA Award for the best Ph.D. thesis in the field of systems and control engineering and he was one of the finalists for the EECI Award for the best Ph.D. thesis in Europe in the field of control for complex and heterogeneous systems.
\end{IEEEbiography}

 \begin{IEEEbiography}[{\includegraphics[width=1in,height=1.25in,clip,keepaspectratio]{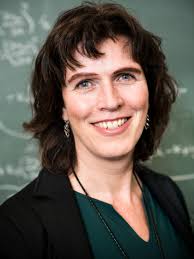}}]{Jacquelien M.A. Scherpen} received the M.Sc.
and Ph.D. degrees in applied mathematics from the
University of Twente, The Netherlands, in 1990 and
1994, respectively. She was with Delft University of
Technology, The Netherlands, from 1994 to 2006.
Since September 2006, she has been a Professor with
the University of Groningen, The Netherlands, at
the Engineering and Technology institute Groningen
(ENTEG) of the Faculty of Science and Engineering.
From 2013 til 2019 she was scientific director of
ENTEG. She is currently Director of the Groningen
Engineering Center. She has held various visiting research positions at
international universities. Her current research interests include model reduction
methods for networks, nonlinear model reduction methods, nonlinear
control methods, modeling and control of physical systems with applications
to electrical circuits, electromechanical systems, mechanical systems, and
grid application, and distributed optimal control applications to smart grids.
Jacquelien Scherpen has been an Associate Editor of the IEEE Transactions on
Automatic Control, the International Journal of Robust and Nonlinear Control
(IJRNC), and the IMA Journal of Mathematical Control and Information. She
is on the Editorial Board of the IJRNC. She was awarded the best paper prize
for Automatica 2017-2020. In 2019 she received a royal distinction and is
appointed Knight in the Order of the Netherlands Lion. Since 2020 she is
Captain of Science of the Dutch topsector High Tech Systems and Materials.
She is council member of IFAC, member of the BoG of the IEEE Control
Systems Society, and president of the European Control Association (EUCA).
\end{IEEEbiography} 
 
\begin{IEEEbiography}[{\includegraphics[trim=40 0 0 0,width=1in,height=1.25in,clip,keepaspectratio]{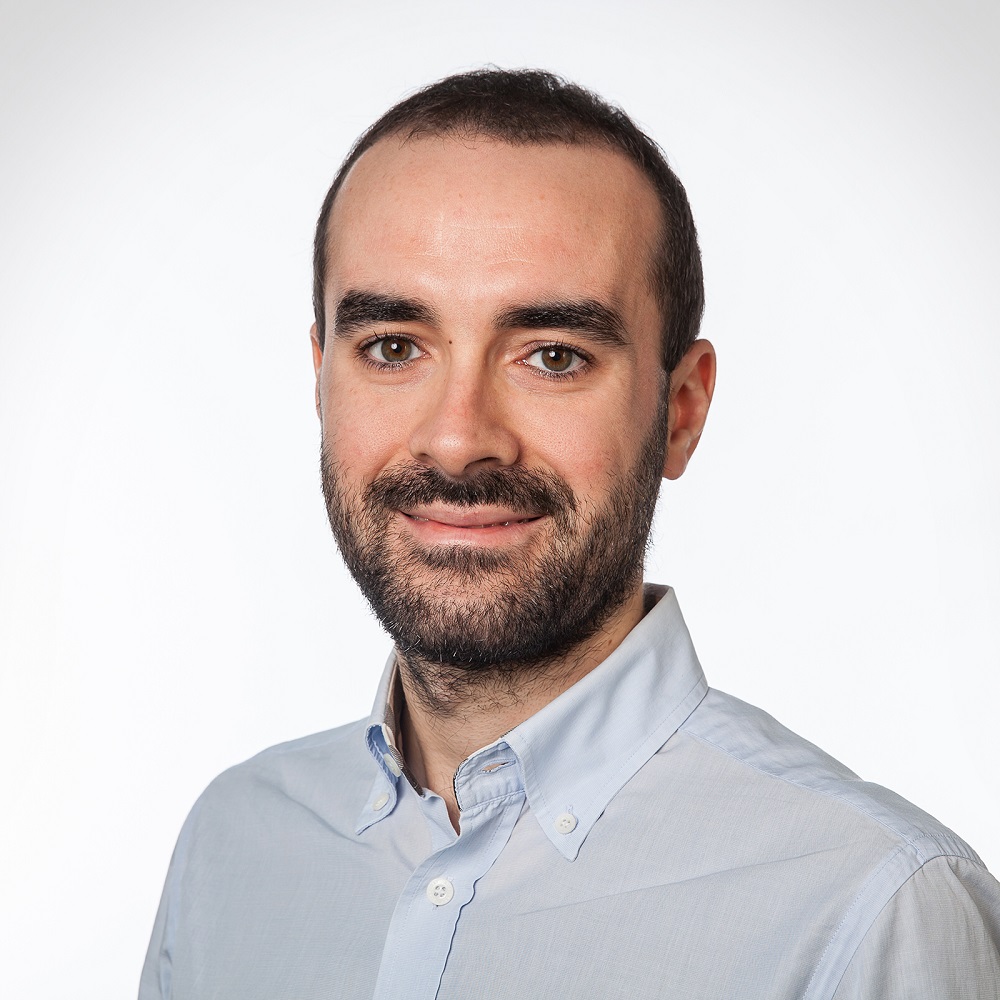}}]{Sergio Grammatico}(M’16 - SM’19) is an associate professor at the Delft Center for Systems and Control, TU Delft, The Netherlands. He received B.Sc., M.Sc. and Ph.D. degrees in Automatic Control Engineering from the University of Pisa, Italy, in 2008, 2009, and 2013 respectively, and an M.Sc. degree in Engineering Science from the Sant’Anna School of Advanced Studies, Pisa, Italy, in 2011. Prior to joining TU Delft, he was a post-doctoral researcher in the Automatic Control Laboratory, ETH Zurich, Switzerland, and an assistant professor in the Department of Electrical Engineering, TU Eindhoven, The Netherlands. He was awarded a 2005 F. Severi B.Sc. scholarship by the Italian High-Mathematics National Institute, and a 2008 M.Sc. fellowship by the Sant’Anna School of Advanced Studies. He was awarded 2013 and 2014 TAC Outstanding Reviewer and he was a recipient of the Best Paper Award at the 2016 ISDG International Conference on Network Games, Control and Optimization. His research interests include distributed optimization and monotone game theory for complex systems of systems. He is currently an Associate Editor for the IEEE Transactions on Automatic Control 34 and Automatica.
\end{IEEEbiography} 

\begin{IEEEbiography}[{\includegraphics[trim=80 0 60 0,width=1in,height=1.25in,clip,keepaspectratio]{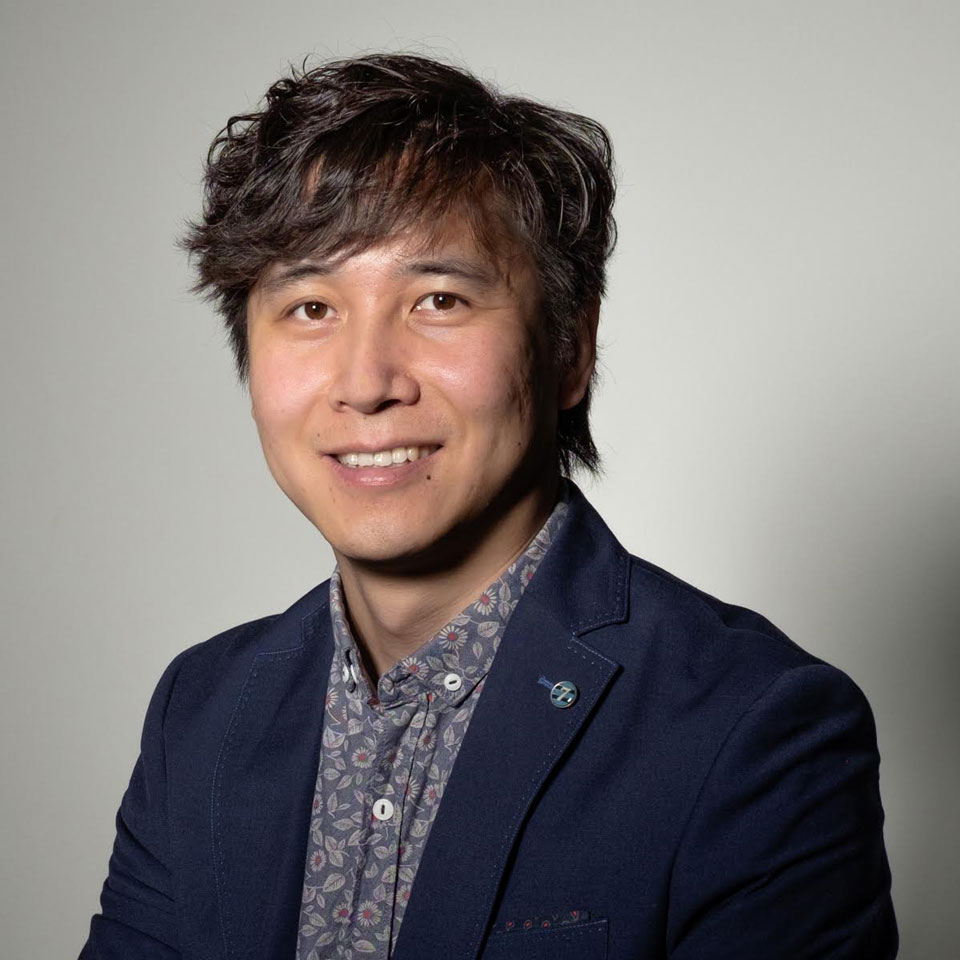}}]{Ming Cao}
has since 2016 been a professor of systems and control with the Engineering and Technology Institute (ENTEG) at the University of Groningen, the Netherlands, where he started as a tenure-track Assistant Professor in 2008. He received the Bachelor degree in 1999 and the Master degree in 2002 from Tsinghua University, Beijing, China, and the Ph.D. degree in 2007 from Yale University, New Haven, CT, USA, all in Electrical Engineering. From September 2007 to August 2008, he was a Postdoctoral Research Associate with the Department of Mechanical and Aerospace Engineering at Princeton University, Princeton, NJ, USA. He worked as a research intern during the summer of 2006 with the Mathematical Sciences Department at the IBM T. J. Watson Research Center, NY, USA. He is the 2017 and inaugural recipient of the Manfred Thoma medal from the International Federation of Automatic Control (IFAC) and the 2016 recipient of the European Control Award sponsored by the European Control Association (EUCA). He is a Senior Editor for Systems and Control Letters, and an Associate Editor for IEEE Transactions on Automatic Control, IEEE Transactions on Circuits and Systems and IEEE Circuits and Systems Magazine. He is a vice chair of the IFAC Technical Committee on Large-Scale Complex Systems. His research interests include autonomous agents and multi-agent systems, complex networks and decision-making processes.  
\end{IEEEbiography}

\end{document}